\documentclass[journal]{IEEEtran}
\usepackage{todonotes} 

\usepackage{tabularx}
\usepackage{amsmath,amssymb,bm,mathrsfs}
\usepackage{siunitx}

\newcommand{\mi}{\textcolor{black}{j}}

\usepackage{breqn}
\usepackage{amsmath}
\usepackage{graphicx}
\usepackage{mwe}
\usepackage{arydshln}
\usepackage{makecell}
\usepackage{adjustbox}
\usepackage{float}
\usepackage{nomencl}
\makenomenclature
\usepackage{mathtools}
\usepackage{bm}
\usepackage{soul}
\usepackage{siunitx}
\usepackage{cite}
\usepackage{multirow}
\usepackage{arydshln}
\usepackage{xfrac}
\usepackage{nicefrac}
\usepackage{nicematrix,tikz}
\usepackage[T1]{fontenc}
\usepackage{inputenc}
\usepackage{babel}
\usepackage[font=small,labelfont=bf]{caption}
\DeclareSIUnit \var { var } 
\usepackage{titlesec}
\usepackage{etoolbox}
\usepackage{arydshln} 
\usepackage{environ}         
\usepackage{etoolbox}        
\usepackage{graphicx}        
\usepackage{hyperref}
\DeclareMathAlphabet\mathbfcal{OMS}{cmsy}{b}{n}
\newtheorem{theorem}{\textbf{Corollary}}

\newlength{\myl}
\let\origequation=\equation
\let\origendequation=\endequation

\RenewEnviron{equation}{
  \settowidth{\myl}{$\BODY$}                       
  \origequation
  \ifdimcomp{\the\linewidth}{>}{\the\myl}
  {\ensuremath{\BODY}}                             
  {\resizebox{0.89\linewidth}{!}{\ensuremath{\BODY}}}  
  \origendequation
}
\newcommand{\overbar}[1]{\mkern 1.5mu\overline{\mkern-1.5mu#1\mkern-1.5mu}\mkern 1.5mu}

\begin{document}

\title{Analytical Computation of the Sensitivity Coefficients
in Hybrid AC/DC Networks \\
\thanks{The project has received funding from the European Unions Horizon 2020 Research \& Innovation Programme under grant agreement No. 957788.}
}

\author{\IEEEauthorblockN{Willem Lambrichts, Mario Paolone\\
Distributed Electrical Systems Laboratory, EPFL, Switzerland \\
willem.lambrichts@epfl.ch, mario.paolone@epfl.ch}
}

\maketitle

\begin{abstract}
In this paper, we present a closed-form model for the analytical computation of the power flow sensitivity coefficients (SCs) for hybrid AC/DC networks. The SCs are defined as the partial derivates of the nodal voltages with respect to the active and reactive power injections. 
The proposed method is inspired by an existing SC computation process proposed for AC networks and here extended to include both the DC grid and the relevant AC/DC Interfacing Converters (ICs). The ICs can operate under different control modes i.e. voltage or power. Additionally, the model is able to compute the SCs for three-phase networks subjected to unbalanced loading conditions. 
The proposed method is numerically validated by means of a comparison with a detailed time-domain simulation model solved within the EMTP-RV simulation environment. Furthermore, we provide a formal proof regarding the uniqueness of the proposed SCs computational model for hybrid AC/DC networks.


\end{abstract}


\section{Introduction}
\label{section:intro}

    
            


Hybrid AC/DC microgrids are a promising solution for future power grids relying heavily on renewable sources. Integrating AC and DC networks has several advantages: 1) an increased overall efficiency of the system, because DC sources and loads are directly connected in the DC grid and thus fewer power conversion sources are required \cite{eghtedarpour2014power}; 2) a lower infrastructure investment cost because of the material savings from cables and transformers and 3) a more flexible grid control that is mainly driven by the controllability of the AC/DC Interfacing Converters (ICs) \cite{BookACDCcontrol, eghtedarpour2014power}.

Grid-aware real-time control is a critical element that is desired for a secure and optimal operation of these hybrid AC/DC networks. One of the main blocks of any real-time control is the Optimal Power Flow (OPF), which aims at computing the optimal setpoints of the controllable Distributed Energy Resources (DER) in both the AC and DC grid and the optimal ICs' setpoints \cite{molzahn2017survey}. 

OPF-type controls require an accurate model of the full hybrid network that is typically defined by the power flow (PF) model. The PF equations for the AC and DC network are strongly non-convex and, therefore, difficult to solve in order to determine the global minimum. Additionally, accurate models of the ICs that link the AC and DC networks need to be included. The inner control loops of the ICs, which are typically operated as Voltage Source Converters (VSC), decouple the d and q frames. Therefore, two electrical quantities can be controlled simultaneously e.g. the DC voltage and the AC reactive power, or the AC active power and the AC reactive power. The control modes are referred to as voltage control: $V_{dc} - Q_{ac}$ and power control: $P_{ac} - Q_{ac}$. As a consequence, the ICs' model has to be suitably adapted depending on the control mode.
Furthermore, the real-time control algorithm typically requires a fast and accurate computation of the OPF problem for time-critical control applications.

Various methods have been presented in the literature for the solution of the OPF problem in hybrid AC/DC networks. Typically, the convexification of the PF and ICs' models is achieved by either relaxing the non-convex constraints or by linearizing them.
Most of the works on hybrid AC/DC OPF are based on the first method where relaxation techniques are deployed. Therefore, new equations are typically added to identify the structure of a new feasible set that is also convex. 
Reference  \cite{wu2021distributed} transforms the non-convex optimisation problem in a semi-definite program (SDP) and shows that the SDP relaxation is exact under specific technical conditions. The authors of \cite{bahrami2016semidefinite} also propose an SDP relaxation to convexify the power losses model and operational constraints of the VSC. 
Reference \cite{baradar2013second}  presents a second-order cone programming (SOCP) relaxation and models the ICs as dummy generators that absorb (or inject) active and reactive power into the AC network. The DC side is modelled similarly and includes a proportional loss. \cite{renedo2019simplified} follows a similar approach. In \cite{li2018optimal} the authors use a SOCP for the OPF in stand-alone DC microgrids. Reference \cite{alvarez2021universal} proposes a method to include the ICs that are able to operate under different operation modes. The method is implemented as an extension of the MATPOWER package \cite{zimmerman2010matpower}. References \cite{cao2013minimization} and \cite{mevsanovic2018robust} solve the OPF problem where the ICs are only operated using droop control. In \cite{hotz2019hynet} an open-source framework for the unified OPF of hybrid High-Voltage DC is presented. The work uses a state space relaxation that relaxes the DC states to voltage phasors and solves the hybrid network as an equivalent AC network.

In the second approach, the grid and ICs models are simplified by, for instance, linearizing the underlying models around their operating point. This approach approximates the model, and thus, may reduce its accuracy. However, because of the model's linear properties, the OPF problem can be solved very efficiently, which is important for real-time optimal control. The method is associated with the concept of sensitivity coefficients (SCs) i.e. the partial derivatives of phase-to-ground voltages (or branch currents) with respect to the control variables (e.g. nodal power injections) \cite{christakou_SC}. The SCs can directly be used in the OPF problem to formulate the grid constraints in a linear way. The closed-loop formulation of the SC requires only the knowledge of the state of the grid and its admittance matrix. When coupled with a state estimation algorithm that provides the grid state in a (sub)second range, the linear approximation is very good because the evaluation of the system states is slow enough compared to the control dynamics. Therefore, the SC-based method is very well suited for real-time control of time-critical applications.
Reference \cite{gupta2019performance} uses SCs of the voltage, current and losses to develop a linear model for AC grids. The authors show that when the SCs are updated dynamically, their use results in a convergence speed and accuracy compatible with real-time control requirements. 
In the application of hybrid AC/DC grids, reference \cite{yang2017optimal} proposes a linear approximation where the square of the DC voltage is used as an independent variable. The ICs' models are approximated using a first-order Taylor expansion. 

The main limitations of the first approach based on the relaxation of the non-convex constraints are the high computation time or the exactness of the solution, which cannot always be guaranteed. Furthermore, none of the existing methods allow more than one IC to regulate the DC voltage. This greatly reduces the redundancy, limits the flexibility of the control and the reduces security of supply, e.g. during islanding manoeuvres. Indeed, the presence of multiple voltage-controlled ICs allows to share the power required to obtain the DC voltage over multiple ICs. This creates a larger solution space and is beneficial in the event of failures. Furthermore, having multiple ICs regulating the DC voltage may result in more realistic grid architectures \cite{barcelos2022direct}.

Therefore, in this work, we propose an accurate and computationally efficient solution for the SCs of hybrid AC/DC networks, inspired by an analytical computation method of the SCs proposed in \cite{christakou_SC}, \cite{christakou2015real}. The authors of \cite{christakou_SC} developed a state-dependent closed-form linear expression that allows for an efficient computation of the voltage SCs. Furthermore, the uniqueness of the SCs is guaranteed.
Here, the closed-form expression for the SCs computation is extended for hybrid AC/DC networks using the unified Power Flow (PF) model of \cite{willem_PF}. The hybrid model includes the AC network, DC network, and ICs, which can operate on different control modes (voltage or power control). Compared to other works presented in the literature, the proposed method allows to consider multiple ICs to regulate the DC voltage.

The paper is structured as follows: Section \ref{section:PFmodel} presents briefly the unified PF model of the hybrid AC/DC network and the nomenclature used throughout this work. Section \ref{section:SCmodel} introduces the closed-form model that allows for the analytical computation of the voltage SCs in hybrid AC/DC networks. In Section \ref{section:Num_validation}, the analytical model is numerically validated.

\section{Unified Power Flow Model for Hybrid AC/DC Networks}
\label{section:PFmodel}

\begin{figure}[!h]
\centering
  \includegraphics[width=0.99\linewidth]{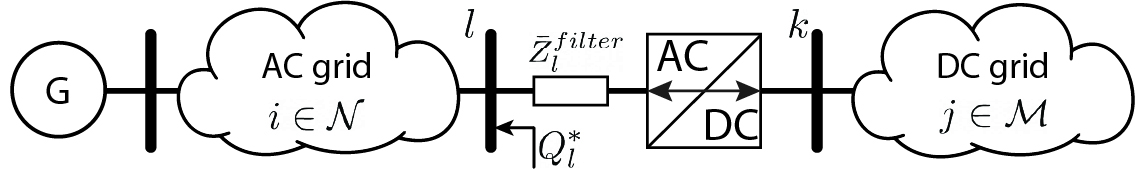}
  \caption{The generic hybrid AC/DC network. Only one AC/DC converter is considered for simplicity.}
  \label{gengrid}
\end{figure}

The closed-form expression of the voltage SC is derived starting from the unified PF model for hybrid AC/DC networks presented in \cite{willem_PF}. We consider a generic hybrid AC/DC grid shown in Figure \ref{gengrid}. The grid consists of $i \in \mathcal{N}$ AC nodes, $j \in \mathcal{M}$ DC nodes and the pair $(l,k) \in \mathcal{L}$ is the couple of AC/DC converter nodes. We assume that $l \in \mathcal{N}$ and $k \in \mathcal{M}$.

The \textbf{AC system} is modelled using the traditional power system theory and consists of three types of nodes: a \textit{Slack} node $(\mathcal{N}_{slack})$, \textit{PQ} nodes $(\mathcal{N}_{PQ})$ and \textit{PV} nodes $(\mathcal{N}_{PV})$. Furthermore, we assume that the zero-injection nodes can be modelled as \textit{PQ} nodes. The AC network is described by its compound three-phase nodal admittance matrix. Therefore, $\overbar{\mathbf{I}}^{ac} = \overbar{\mathbf{Y}}^{ac} \overbar{\mathbf{E}}^{ac} $, where $\overbar{\mathbf{E}}^{ac}$ is the phase-to-ground nodal voltage vector, $\overbar{\mathbf{I}}^{ac}$ the nodal current injections and $\overbar{\mathbf{Y}}^{ac}$ the compound admittance matrix, which is assumed to be known. 

The \textbf{DC system} is modelled identically to the AC network using the classic AC theory where the electrical quantities are strictly real values: $Q = 0$ and $\overbar{Z} = R$. Two types of controllable nodes are introduced: constant power nodes $(\mathcal{M_{P}})$ and constant voltage nodes $(\mathcal{M_{V}})$. The DC network is described by its compound admittance matrix. Therefore, ${\mathbf{I}}^{dc} = {\mathbf{Y}}^{dc} {\mathbf{E}}^{dc} $, where ${\mathbf{E}}^{dc}$ is the phase-to-ground nodal voltage vector, ${\mathbf{I}}^{ac}$ the nodal current injections and ${\mathbf{Y}}^{ac}$ the compound admittance matrix, which is again assumed to be known.

The \textbf{interfacing converters}, typically operated as VSCs, interface the AC and the DC system in one or more nodes (i.e., $\lvert \mathcal{L} \rvert \geq 1$). The control scheme allows the VSC to operate in different modes and control different electrical quantities. Usually, the control pairs are the active and reactive power injection: $P_{ac}-Q_{ac}$, or the DC voltage and the reactive power: $V_{dc}-Q_{ac}$. The ICs use a phase-locked loop (PLL) based control scheme to synchronise to the grid. During unbalanced loading conditions, the PLL typically synchronises with the positive sequence components to only inject positive sequence power, i.e., the homopolar and negative sequence components of the injected power are zero. In specific cases, the ICs can also intentionally inject negative sequence power to reduce the unbalanced loading conditions.  
    
In conclusion, the set of AC and DC nodes is described as:
\begin{align}
    &\mathcal{N} = \mathcal{N}_{slack} \cup \mathcal{N}_{PQ} \cup \mathcal{N}_{PV} \cup \mathcal{L}_{l,  PQ} \cup \mathcal{L}_{l,  V_{dc}Q}, \nonumber \\
    &\mathcal{M} = \mathcal{M}_{P_{dc}} \cup \mathcal{M}_{V_{dc}} \cup \mathcal{L}_{k, PQ} \cup \mathcal{L}_{k, V_{dc}Q}, \nonumber
\end{align}
and the different nodes in the hybrid AC/DC network are summarised in Table \ref{NodeTypes}. The nomenclature and indices of the node types of Table  \ref{NodeTypes} are used consistently throughout this paper.

\normalsize
\renewcommand{\arraystretch}{1.2}
\begin{table}[!h]
\caption{Different types of nodes in hybrid AC/DC networks and their known and unknown variables.}
\resizebox{1\columnwidth}{!}{%
\begin{tabular}{lllll}
\hline
Bus Type                & IC contrl.                      & Known var.                                              & Unknown var.  & Index\\ \hline
AC slack                   &                                  & $\lvert E_{ac} \rvert$, $\angle E_{ac}$                               & $P_{ac}$,$Q_{ac}$       & $s \in \mathcal{N}_{slack}$   \\ \hline
$P_{ac}$, $Q_{ac}$                 &                                  & $P_{ac}$,$Q_{ac}$                                                & $\lvert E_{ac} \rvert$, $\angle E_{ac}$    & $i \in \mathcal{N}_{PQ}$            \\ \hline
$P_{ac}$, $\lvert V_{ac} \rvert$                 &                                  & $P_{ac}$,$\lvert V_{ac} \rvert$                                                & $Q_{ac} $, $\angle E_{ac}$  & $i \in \mathcal{N}_{PV}$              \\ \hline
\multirow{2}{*}{$IC_{ac}$} & $P$ - $Q $                & $P_{ac}$ $Q_{ac}$     & $\lvert E_{ac} \rvert$, $\angle E_{ac}$           &  $l \in \mathcal{L}_{PQ}$  \\ \cdashline{2-5}
                        & $V_{dc}$ - $Q  $                   & $Q_{ac}$              & $P_{ac}$, $\lvert E_{ac} \rvert$, $\angle E_{ac}$     &  $l \in \mathcal{L}_{V_{dc}Q}$        \\ \hline
\multirow{2}{*}{$IC_{dc}$} & $P$ - $Q$                 & $P_{dc}$              & $E_{dc}$           &  $k \in \mathcal{L}_{PQ}$   \\ \cdashline{2-5}
                        & $V_{dc}$ - $Q $                    & $E_{dc}$              & $P_{dc}$           &   $k \in \mathcal{L}_{V_{dc}Q}$  \\  \hline
$P_{dc}$                     &                                  & $P_{dc}$                                                    & $E_{dc}$          & $j \in \mathcal{M}_{P}$    \\ \hline
$E_{dc}$                     &                                  & $E_{dc}$                                                    & $P_{dc}$           & $j \in \mathcal{M}_{V}$   \\ \hline
\end{tabular} \label{NodeTypes}
}
\end{table}
\normalsize

It is worth noticing that the DC grid doesn't have a 'real' DC slack bus. However, in at least one of the nodes, the voltage needs to be regulated to ensure a safe operation of the DC grid, this can be either an IC operating in $V_{dc}- Q$ mode or a DC voltage node $\mathcal{M}_{V_{dc}}$.

The generic PF equations for a hybrid, unbalanced AC/DC network are shown in \eqref{eq:PFmodel}. The model allows for multiple ICs with different operating modes \cite{willem_PF}. Equations \ref{PQ_p} - \ref{PV} model the \textit{PQ} and \textit{PV} nodes in the AC grid, equations \ref{P_dc} and \ref{V_dc} model the \textit{P} and \textit{V} nodes in DC network and equations \ref{EP} - \ref{E-} model the different operating modes of the ICs.

\begin{subequations}
\small
\allowdisplaybreaks
\begin{align}
&\text{\textbf{AC nodes}} \nonumber \\
        &\Re \Big\{ \overbar{E}_{i}^{\phi} \sum\nolimits_{n \in \mathcal{N}} \underbar Y_{i,n}^{ac} \underbar E_{n}^{\phi} \Big\} = P^{\phi \ast}_{i} , &&  \forall i \in \mathcal{N}_{PQ} \cup \mathcal{N}_{PV} 
        \label{PQ_p}\\ 
        &\Im \Big\{ \overbar{E}_{i}^{\phi} \sum\nolimits_{n \in \mathcal{N}} \underbar Y_{i,n}^{ac} \underbar E_{n}^{\phi} \Big\} = Q^{\phi \ast}_{i} , &&  \forall i \in \mathcal{N}_{PQ} \label{PQ_q}\\ 
        & \big( {E_{i}^{\phi \prime}} \big) ^2 + \big( {E_{i}^{\phi \prime \prime}} \big)^2 = \lvert {E_{i}^{\phi \ast}} \rvert ^2 , 
        &&  \forall i \in \mathcal{N}_{PV} \label{PV} \\
        \nonumber \\
&\text{\textbf{DC nodes}}  \nonumber \\
        & E_{j} \sum\nolimits_{m \in \mathcal{M}} Y_{j,m}^{dc} E_{m}  = P^{\ast}_{j} , 
        && \forall j \in \mathcal{M}_{P} \label{P_dc}\\ 
        & E_{j} = E^{\ast}_{j} , 
        && \forall j \in \mathcal{M}_{V} \label{V_dc} \\
        \nonumber \\
&\text{\textbf{IC nodes}} \nonumber \\
       & E_{k}^{\ast} \Big( Y_{(k,k)}^{dc} E_{k}^{\ast} + \sum\nolimits_{\substack{m \in \mathcal{M} \\
              m \neq k}} Y_{(k,m)}^{dc} E_{m} \Big)  
              && \nonumber \\
       &  \qquad \quad  = \Re  \Big\{ \overbar{E}_{l}^+ \sum\nolimits_{n \in \mathcal{N}} \underbar Y_{(l,n)}^{ac} \underbar E_{n}^+ \Big\},  
       && \forall (l,k) \in \mathcal{L}_{V_{dc}Q}   \label{Edc}\\
        &  \Re \Big\{ \overbar{E}_{l}^+ \sum\nolimits_{n \in \mathcal{N}} \underbar Y_{(l,n)}^{+ ac} \underbar E_{n}^+ \Big\}   = P^{\ast}_l  ,  
        && \forall l \in \mathcal{L}_{PQ} 
        \label{EP} \\ 
       &  \Im \Big\{ \overbar{E}_{l}^+ \sum\nolimits_{n \in \mathcal{N}} \underbar Y_{(l,n)}^{+, ac} \underbar E_{n}^+ \Big\}   = Q^{\ast}_l ,  &&  \forall l \in \mathcal{L}_{PQ} \cup \mathcal{L}_{V_{dc}Q} 
       \label{EQ} \\
        &E^{0  \prime}_l  = 0 , \qquad \ E^{0  \prime \prime}_l  = 0 , && \forall l \in \mathcal{L}_{PQ} \cup \mathcal{L}_{V_{dc}Q} 
        \label{E0} \\ 
        &E^{n  \prime}_l  = 0 , \qquad E^{n  \prime \prime}_l  = 0,  && \forall l \in \mathcal{L}_{PQ} \cup \mathcal{L}_{V_{dc}Q} 
        \label{E-} 
\end{align} \label{eq:PFmodel}
\end{subequations}

Where, $\phi$ is the phase: $\phi \in \{a,b,c \}$, the underbar $\underbar{$\square$}$ indicates the complex conjugate and the prime symbols $ \square^{\prime} $ and $ \square^{\prime \prime} $ 
refer to the real and imaginary parts of the complex electrical quantities. The subscripts {\small $0, +, -$} denote the homopolar (zero), positive and negative sequence components following the standard symmetrical component decomposition and the asterisks $\square^{\ast}$ refers to the controllable variable.


\section{Methodology}
\label{section:SCmodel}


Using the power flow model in \eqref{eq:PFmodel} a closed-form mathematical expression is derived to compute the voltage SC for the AC nodes, DC nodes and IC nodes in a unified way. In order not to lose the generality of the method, the voltage SC are computed with respect to the set $\mathcal{X}$ independent variables of the PF model \eqref{eq:PFmodel}. Furthermore, let $x$ be any element from the set $\mathcal{X}$. 
Therefore,
\begin{align}
&\mathcal{X} = \left\{ P^{\phi \ast}_{i},Q^{\phi \ast}_{i}, \lvert \overbar{E}_{i}^{\phi \ast} \rvert ,P^{\ast}_{j},E^{\ast}_{j},P^{\ast}_l,Q^{\ast}_l , E_{k}^{\ast} \right\} \nonumber \\ 
& \qquad \qquad \qquad \qquad \qquad \forall \ i \in \mathcal{N}, \forall \ j \in \mathcal{M}, \forall \ (l,k) \in \mathcal{L}  \label{eq:X}
\end{align}

\noindent
In what follows, we derive the SCs with respect to each element in $\mathcal{X}$ using the procedure: 
\begin{enumerate}
    \item Compute the partial derivative with respect to $x \in \mathcal{X}$ of the PF equations \eqref{eq:PFmodel}. 
    \item Regroup the partial derivatives to obtain the linear system of equations $\mathbf{A} \  \mathbf{u}(x) = \mathbf{b}(x)$, where $\mathbf{u}(x)$ is the vector of the voltage SCs $\frac{\partial \overbar{E}}{\partial x}$ as defined in \eqref{d_dX}. 
    \small
\begin{align}
&\mathbf{u}(x) =  \left[ \frac{\partial \overbar{E}_i^{\phi '}}{ \partial x}, \frac{\partial \overbar{E}_i^{\phi ''}}{ \partial x},
                                    \frac{\partial {E}_j}{ \partial x} , 
                                    \frac{ \partial \overbar{E}_l^{\phi '}}{ \partial x}, \frac{ \partial \overbar{E}_l^{\phi ''}}{ \partial x} ,
                                    \frac{ \partial {E}_k}{ \partial x}
                            \right] \nonumber \\ 
&  \forall \ i \in \mathcal{N},  \ j \in \mathcal{M},  \ (l,k) \in \mathcal{L} , \ \phi \in \{a,b,c \},  \ x \in \mathcal{X} \label{d_dX}
\end{align} \normalsize
    \item Solve the linear system of equations to obtain $\mathbf{u}(x)$.
\end{enumerate} 

\noindent
The assumptions made in the closed-form computation of the voltage SCs are reported here below.
    \begin{itemize}

    \item The voltage of the \textit{Slack} bus $\overbar{E}_s^{\phi}$ is fixed and can therefore, not be influenced by any control variables other than itself. Therefore,
    \begin{align}
    &\frac{\partial \lvert \overbar{E}_s^{\phi '} \rvert }{ \partial \lvert \overbar{E}_s^{\phi '} \rvert }  = 1, \frac{\partial \lvert \overbar{E}_s^{\phi ''} \rvert }{ \partial \lvert \overbar{E}_s^{\phi ''} \rvert }  = 1, \frac{\partial \lvert \overbar{E}_s^{\phi '} \rvert }{ \partial x }  = 0, \frac{\partial \lvert \overbar{E}_s^{\phi ''} \rvert }{ \partial x }  = 0, \nonumber \\
    & \qquad \qquad \qquad \qquad \qquad \qquad \qquad \qquad \forall \ x \in \mathcal{X}
    \end{align} \normalsize
    The voltage SC of the slack node can be computed directly and is not required to be included in the linear system of equations. Therefore, the slack voltage is also not included in the set of controllable variables $\mathcal{X}$.

    \item Trivially, the partial derivative of any voltage with respect to itself is 1. This is relevant for the voltage-controllable nodes: the \textit{PV} nodes in the AC grid, \textit{V} nodes in the DC grid and the IC operating under $V_{dc} - Q$ mode. 
    \small
    \begin{align}
    &\frac{\partial \lvert \overbar{E}_i^{\phi \ast} \rvert }{ \partial \lvert \overbar{E}_i^{\phi \ast} \rvert }  = 1, && \qquad \forall i \in \mathcal{N}_{PV}, \nonumber \\
    &\ \ \frac{\partial  {E}_j^{\ast}  }{ \partial  {E}_j^{\ast}  }  \ = 1, && \qquad \forall j \in \mathcal{M}_{V}, \nonumber \\
    &\ \ \frac{\partial  {E}_k^{\ast}  }{ \partial  {E}_k^{\ast}  }  \ = 1, && \qquad \forall k \in \mathcal{L}_{V_{dc}Q} \label{eq:rules_1}
    \end{align} \normalsize

    \item We assume that the nodal voltage magnitudes of the voltage-controllable AC, DC and IC nodes are fixed and cannot be influenced by any other control variable. Therefore, the partial derivative of these voltages with respect to the other control variables equals zero.
    \small
    \begin{align}
    &\frac{\partial \lvert \overbar{E}_i^{\phi \ast} \rvert }{ \partial x }  = 0,&& \qquad \forall i \in \mathcal{N}_{PV}, \ \forall x \in \mathcal{X} \setminus \{ \lvert \overbar{E}_{i}^{\phi \ast} \rvert \}  \nonumber \\
    &\ \ \frac{\partial  {E}_j^{\ast}  }{ \partial  x } \  = 0, && \qquad \forall j \in \mathcal{M}_{V}, \ \forall x \in \mathcal{X} \setminus \{ {E}_j^{\ast} \}  \nonumber \\
    &\ \ \frac{\partial  {E}_k^{\ast}  }{ \partial  x } \  = 0,&& \qquad \forall k \in \mathcal{L}_{V_{dc}Q}, \ \forall x \in \mathcal{X} \setminus \{ {E}_k^{\ast} \} \label{eq:rules_2}
    \end{align} \normalsize
    
    \end{itemize}

\noindent
Finally, the SC of a voltage in rectangular coordinates can be easily transformed to polar coordinates using the transformations in \eqref{SCidentity2} and  \eqref{SCidentity}. In what follows, the SC will be computed in the rectangular coordinate system.
    \begin{align}
        & \frac{\partial \overbar{E}}{ \partial x}  =  \frac{\partial \overbar{E}^{'}}{ \partial x}  + j  \frac{\partial \overbar{E}^{''}}{ \partial x}  = \overbar{E} \left( \frac{1}{\lvert \overbar{E} \rvert} \frac{\partial \lvert \overbar{E} \rvert}{\partial x} + j \frac{\partial \angle \overbar{E}}{\partial x}
        \right)  \nonumber \\
        & \frac{\partial \underbar{$E$}}{ \partial x}  =  \frac{\partial \overbar{E}^{'}}{ \partial x}  - j  \frac{\partial \overbar{E}^{''}}{ \partial x}  = \underbar{$E$} \left( \frac{1}{\lvert \overbar{E} \rvert} \frac{\partial \lvert \overbar{E} \rvert}{\partial x} - j \frac{\partial \angle \overbar{E}}{\partial x}
        \right)   \label{SCidentity2}
    \end{align}
  and,  
    \begin{align}
    \frac{\partial \lvert \overbar{E} \rvert}{\partial x} = \frac{1}{\lvert \overbar{E} \rvert} \Re \left\{ \underbar{$E$} \frac{\partial \overbar{E}}{ \partial x}  \right\} \
        \text{and} \ \frac{\partial \angle \overbar{E}}{\partial x} = \frac{1}{\lvert \overbar{E} \rvert^2 } \Im \left\{ \underbar{$E$} \frac{\partial \overbar{E}}{ \partial x}  \right\}
        \label{SCidentity}
    \end{align} 
    
\noindent    
Furthermore, to improve the clarity of this paper, the expressions $\phi \in \{a,b,c \}$ and $x \in \mathcal{X}$ are omitted from every equation below, but are valid.


\subsection{Voltage SC in the AC grid}

In the generic case, the voltage SC for the \textit{PQ} and \textit{PV} nodes in the AC grid with respect to $x$ are computed starting from the power flow equation \eqref{PQ_p} and \eqref{PQ_q} that relate the power injection to the nodal voltages.
\begin{align}
    &P^{\phi \ast}_{i} + j Q^{\phi \ast}_{i} = \overbar{E}_{i}^{\phi} {\textstyle\sum\limits_{n \in \mathcal{N}}} \underbar Y_{i,n}^{ac} \underbar E_{n}^{\phi}, &&  \forall i \in \mathcal{N}_{PQ} \label{eq:SC_1}
\end{align}
\normalsize
\noindent
Next, we take the partial derivative of \eqref{eq:SC_1} with respect to $x$. 
\small 
\begin{align} 
    &\frac{\partial P_{i}^{\phi \ast}}{ \partial x} + j \frac{\partial Q_{i}^{\phi \ast}}{ \partial x} = \overbar{E}_{i}^{\phi} \frac{\partial  }{ \partial x} \left(  {\textstyle\sum\limits_{n \in \mathcal{N}}} \underbar Y_{i,n}^{ac} \underbar E_{n}^{\phi} \right)  +  \frac{ \partial \overbar{E}_{i}^{\phi}}{ \partial x}  {\textstyle\sum\limits_{n \in \mathcal{N}}} \underbar Y_{i,n}^{ac} \underbar E_{n}^{\phi}  \nonumber \\
    &\qquad \qquad \qquad \qquad \qquad \qquad \qquad \qquad \qquad \quad \forall i \in \mathcal{N}_{PQ}
     \label{eq:SC_2}
\end{align}
\normalsize

\noindent
Using the identities in \eqref{SCidentity2}, we can reformulate \eqref{eq:SC_2} into its real and imaginary components. Furthermore, because $\overbar{E}_{i}^{\phi}$ in the first term is not dependent on $n$, we can bring it into the summation.

\begin{align} 
    &\frac{\partial P_{i}^{\phi \ast}}{ \partial x} + j \frac{\partial Q_{i}^{\phi \ast}}{ \partial x} = {\textstyle\sum\limits_{n \in \mathcal{N}}} \overbar{E}_{i}^{\phi}  \underbar Y_{i,n}^{ac} \Big( \frac{\partial \overbar E_{n}^{\phi'}  }{ \partial x} - j \frac{\partial \overbar E_{n}^{\phi''}  }{ \partial x}  \Big) \qquad \qquad \nonumber \\
    & \qquad \qquad \qquad \quad \ +  \Big( \frac{\partial \overbar E_{i}^{\phi'}  }{ \partial x} - j \frac{\partial \overbar E_{i}^{\phi''}  }{ \partial x}  \Big)  {\textstyle\sum\limits_{n \in \mathcal{N}}} \underbar Y_{i,n}^{ac} \underbar E_{n}^{\phi} \label{eq:SC_4}
\end{align}
\normalsize

\noindent
Expression  \eqref{eq:SC_4} can be simplified by substituting:
\begin{align}
\overbar{F}_{i,n}^{\phi} = \overbar{E}_{i}^{\phi}  \underbar Y_{i,n}^{ac} \ \ \ \text{\normalsize and} \ \ \ \overbar{H}_{i}^{\phi} = {\textstyle\sum\limits_{n \in \mathcal{N}}} \underbar Y_{i,n}^{ac} \underbar E_{n}^{\phi}    \label{eq:SC_6}
\end{align}
\normalsize

\noindent
Substituting \eqref{eq:SC_6} in \eqref{eq:SC_4} and rearranging the real and imaginary terms, gives the expression for the active and reactive power injections that is linear in $\frac{\partial E}{\partial x}$.

\begin{align} 
    &\frac{\partial P_{i}^{\phi \ast}}{ \partial x}  =
    \left( \overbar{F}_{i,i}^{\phi '} + \overbar{H}_{i}^{\phi '} \right) \frac{\partial \overbar E_{i}^{\phi'}  }{ \partial x}
    \ +  {\textstyle\sum\limits_{\substack{n \in \mathcal{N} \setminus \{i\} }}} \overbar{F}_{i,n}^{\phi '}  \frac{\partial \overbar E_{n}^{\phi'}  }{ \partial x} \nonumber \\
    &\qquad  \ \ + \left( \overbar{F}_{i,i}^{\phi ''}  - \overbar{H}_{i}^{\phi ''}\right)  \frac{\partial \overbar E_{i}^{\phi''}  }{ \partial x} 
    +  {\textstyle\sum\limits_{\substack{n \in \mathcal{N} \setminus \{i\} }}} \overbar{F}_{i,n}^{\phi ''}  \frac{\partial \overbar E_{n}^{\phi''}  }{ \partial x} \nonumber \\
    &  \qquad \qquad \qquad \qquad \qquad \qquad \qquad \quad \forall i \in \mathcal{N}_{PQ} \cup \mathcal{N}_{PV} \label{eq:SC_7a} \\
    &\frac{\partial Q_{i}^{\phi \ast}}{ \partial x}  =
    \left( \ \ \overbar{F}_{i,i}^{\phi ''} + \overbar{H}_{i}^{\phi ''} \right) \frac{\partial \overbar E_{i}^{\phi'}  }{ \partial x}
    +  {\textstyle\sum\limits_{\substack{n \in \mathcal{N} \setminus \{i\} }}} \overbar{F}_{i,n}^{\phi '}  \frac{\partial \overbar E_{n}^{\phi'}  }{ \partial x} \nonumber \\
    &\qquad  \ \ + \left( -\overbar{F}_{i,i}^{\phi '}  + \overbar{H}_{i}^{\phi '}\right)  \frac{\partial \overbar E_{i}^{\phi''}  }{ \partial x} 
    -  {\textstyle\sum\limits_{\substack{n \in \mathcal{N} \setminus \{i\} }}} \overbar{F}_{i,n}^{\phi ''}  \frac{\partial \overbar E_{n}^{\phi''}  }{ \partial x} \nonumber \\
    & \qquad \qquad \qquad \qquad \qquad \qquad \qquad \qquad \forall i \in \mathcal{N}_{PQ}
    \label{eq:SC_7b}
\end{align}
\normalsize

The voltage SCs of the PV nodes in the AC grid are computed starting from the PF model \eqref{PV} that relates the voltage magnitude to its real and imaginary part.
\begin{align}
    \lvert \overbar{E}_{i}^{\phi \ast} \rvert^2 = \big( \overbar{E}_{i}^{\phi '} \big)^2 + \big( \overbar{E}_{i}^{\phi ''} \big)^2,  \quad \forall i \in \mathcal{N}_{PV}
    \label{eq:SC_8}
\end{align}

\noindent
Next, we take the partial derivative of \eqref{eq:SC_8} with respect to $x$ to obtain the linear expression of the voltage sensitivity coefficients.
\begin{align}
    & \overbar{E}_{i}^{\phi \ast} \frac{\partial \lvert \overbar{E}_{i}^{\phi \ast} \rvert }{\partial x} = \overbar{E}_{i}^{\phi '} \frac{\partial \overbar{E}_{i}^{\phi ' }}{\partial x}  +  \overbar{E}_{i}^{\phi ''} \frac{\partial \overbar{E}_{i}^{\phi ''}}{\partial x},  \  \forall i \in \mathcal{N}_{PV}
    \label{eq:SC_9}
\end{align}

\subsection{Voltage SC in the DC grid}

The DC grid model consists of \textit{P} nodes ($ \mathcal{M}_P$) and \textit{V} nodes ($ \mathcal{M}_V$). Because of its DC nature, all the electrical quantities are real: $\overbar{E}_{j} = E_{j}, \ \forall j \in \mathcal{M}$.

The power injection in the DC node $j$ is expressed as presented in the PF model \eqref{P_dc},
\begin{align}
    P^{\ast}_{j} = {E}_{j} {\textstyle\sum\limits_{m \in \mathcal{M}}} Y_{j,m}^{dc} E_{m}, \qquad \qquad \qquad   \forall j \in \mathcal{M}_{P} \label{eq:SCdc_1}
\end{align}

\noindent
The voltage SCs are computed by taking the partial derivative of equation \eqref{eq:SCdc_1} to $x$.

\begin{align} 
    &\frac{\partial P_{j}^{\ast}}{ \partial x} = {E}_{j} {\textstyle\sum\limits_{m \in \mathcal{M}}}  Y_{j,m}^{dc} \frac{\partial  E_{m}  }{ \partial x}  +  \frac{\partial  E_{j} }{ \partial x}  {\textstyle\sum\limits_{m \in \mathcal{M}}}  Y_{j,m}^{dc} E_{m} \nonumber \\
    & \qquad \qquad \qquad \qquad \qquad \qquad \quad  \forall j \in \mathcal{M}_{P} \label{eq:SCdc_3}
\end{align}
\normalsize

\noindent
Using the identities in \eqref{eq:SCdc_4}, the expression \eqref{eq:SCdc_3} is simplified to \eqref{eq:SCdc_5}
\begin{align}
{F}_{j,m} = {E}_{j}  Y_{j,m}^{dc} \ \ \ \text{\normalsize and} \ \ \ {H}_{j} = {\textstyle\sum\limits_{m \in \mathcal{M}}}  Y_{j,m}^{dc}  E_{m}    \label{eq:SCdc_4}
\end{align}

\begin{align}
    &\frac{\partial P_{j}^{\ast}}{ \partial x}  =
    \left( {F}_{j,j} + {H}_{j} \right) \frac{\partial  E_{j} }{ \partial x}
    \ +  {\textstyle\sum\limits_{\substack{m \in \mathcal{M} \setminus \{j\} }}} {F}_{j,m}  \frac{\partial  E_{m}  }{ \partial x} \nonumber \\
    & \qquad \qquad \qquad \qquad \qquad \qquad \quad  \forall j \in \mathcal{M}_{P} \label{eq:SCdc_5}
\end{align}

The voltage SC of the \textit{V} is formulated in \eqref{eq:SCdc_6}.
\begin{align}
        &\frac{\partial E^{\ast}_{j}}{\partial x}  =  \frac{\partial E_j}{\partial x},  
        \qquad \qquad \ \  \forall j \in \mathcal{M}_{V} \label{eq:SCdc_6}
\end{align}

\subsection{Voltage SC for the Interfacing Converters}

Finally, the expressions of the partial derivatives of the IC's voltages have to be formulated to be included in the unified closed-form SC model. Every IC is connected to a node pair $(k,l) \in \mathcal{L}$, where $l$ is the IC's AC node and $k$ is the IC's DC node. To improve the clarity of the model, first, the system of equations is derived for balanced grid conditions. In the next section \eqref{subsec:unbalanced}, the SC model is adapted for unbalanced grid conditions.

For the IC's operating mode $\mathbf{V_{dc} - Q}$, the SCs are computed starting from the PF equations described in \eqref{Edc} and \eqref{EQ}. The model relating the controllable DC voltage to the other AC and DC nodal voltages is given in \eqref{eq:SCic_1}.
\small
\begin{align}
        & \Re  \Big\{ \overbar{E}_{l}^{\phi} {\textstyle\sum\limits_{\substack{n \in \mathcal{N} }}} \underbar Y_{(l,n)}^{ac} \underbar E_{n} \Big\}  = 
         E_{k}^{\phi \ast} \Big( Y_{(k,k)}^{dc} E_{k}^{\ast} + {\textstyle\sum\limits_{\substack{m \in \mathcal{M} \setminus \{k\} }}} Y_{(k,m)}^{dc} E_{m} \Big)  \nonumber \\ 
        &  \qquad \qquad  \qquad \qquad \qquad \qquad  \qquad  \qquad \quad   \forall (l,k) \in \mathcal{L}_{V_{dc}Q}  \label{eq:SCic_1}
\end{align}
\normalsize

\noindent
Next, we take the partial derivative of \eqref{eq:SCic_1} to $x$. The derivative of the left-hand side of \eqref{eq:SCic_1} is computed analogously to \eqref{eq:SC_6}, however, the indices have to be suitably adapted to the IC's AC node $l$. The derivative of the right-hand side reads:
\begin{align}
        & \frac{\partial E_{k}^{\ast}  }{ \partial x} \Big( Y_{(k,k)}^{dc} E_{k}^{\ast} + {\textstyle\sum\limits_{\substack{m \in \mathcal{M} \setminus \{k\} }}}  Y_{(k,m)}^{dc} E_{m} \Big) + \nonumber \\
        & \ \ E_{k}^{\ast} \Big( Y_{(k,k)}^{dc} \frac{\partial E_{k}^{\ast}  }{ \partial x} + {\textstyle\sum\limits_{\substack{m \in \mathcal{M} \setminus \{k\} }}}  Y_{(k,m)}^{dc} \frac{\partial  E_{m}  }{ \partial x} \Big)
\end{align}

\noindent
Finally, we regroup the terms to obtain a linear relation of the partial derivative of the controllable DC voltage $\frac{\partial E_{k}^{\ast}  }{ \partial x}$. Furthermore, the expression is simplification using \eqref{eq:SCdc_4}.
\begin{align}
    & \Big( F_{(k,k)} + H_{k} \Big) \frac{\partial E_{k}^{\ast}  }{ \partial x}  = -{\textstyle\sum\limits_{\substack{m \in \mathcal{M} \setminus \{k\} }}}  F_{(k,m)} \frac{\partial  E_{m}  }{ \partial x} \nonumber \\
    & \qquad + \left( \overbar{F}_{l,l}^{\phi '} + \overbar{H}_{l}^{\phi '} \right) \frac{\partial \overbar E_{l}^{\phi'}  }{ \partial x}
    \ +  {\textstyle\sum\limits_{\substack{n \in \mathcal{N} \setminus \{l\} }}} \overbar{F}_{l,n}^{\phi '}  \frac{\partial \overbar E_{n}^{\phi'}  }{ \partial x} \nonumber \\
    & \qquad + \left( \overbar{F}_{l,l}^{\phi ''}  - \overbar{H}_{l}^{\phi ''}\right)  \frac{\partial \overbar E_{l}^{\phi''}  }{ \partial x}  +  {\textstyle\sum\limits_{\substack{n \in \mathcal{N} \setminus \{l\} }}} \overbar{F}_{i,n}^{\phi ''}  \frac{\partial \overbar E_{n}^{\phi''}  }{ \partial x} \nonumber \\
    & \qquad \qquad  \qquad \qquad \qquad \qquad \qquad  \forall (l,k) \in \mathcal{L}_{V_{dc}Q} \label{eq:SCic_v}
\end{align}

For the operating mode $\mathbf{P-Q}$, the partial derivatives of the active power injections of the IC into the AC grid are described similarly to \eqref{eq:SC_7a}.
\begin{align}
&\frac{\partial P_{l}^{\phi \ast}}{ \partial x}  =
    \left( \overbar{F}_{l,l}^{\phi '} + \overbar{H}_{l}^{\phi '} \right) \frac{\partial \overbar E_{l}^{\phi'}  }{ \partial x}
    \ +  {\textstyle\sum\limits_{\substack{n \in \mathcal{N} \setminus \{i\} }}} \overbar{F}_{l,n}^{\phi '}  \frac{\partial \overbar E_{n}^{\phi'}  }{ \partial x} \nonumber \\
    &\qquad  \ \ + \left( \overbar{F}_{l,l}^{\phi ''}  - \overbar{H}_{l}^{\phi ''}\right)  \frac{\partial \overbar E_{l}^{\phi''}  }{ \partial x} 
    +  {\textstyle\sum\limits_{\substack{n \in \mathcal{N} \setminus \{l\} }}} \overbar{F}_{l,n}^{\phi ''}  \frac{\partial \overbar E_{n}^{\phi''}  }{ \partial x} \nonumber \\
    &  \qquad \qquad  \qquad \qquad \qquad \qquad \qquad \forall l \in \mathcal{L}_{PQ}  \label{eq:SCic_p}
\end{align}

Typically, the IC operating in $\mathbf{P-Q}$ mode, are given an AC power reference to track. Using the active power balance at the IC between the AC and the DC network, we can say that $P_{l}^{\phi \ast} = P_{k}^{\ast}$. Therefore, the partial derivatives of the active power injected into the DC grid is described in \eqref{eq:SCic_pdc}
\begin{align}
    &\frac{\partial P_{k}^{\ast}}{ \partial x}  =
    \left( {F}_{k,k} + {H}_{k} \right) \frac{\partial  E_{k} }{ \partial x}
    \ +  {\textstyle\sum\limits_{\substack{m \in \mathcal{M} \setminus \{k\} }}} {F}_{k,m}  \frac{\partial  E_{m}  }{ \partial x} \nonumber \\
    & \qquad \qquad \qquad \qquad \qquad \qquad \qquad \quad  \forall k \in \mathcal{L}_{V_{dc}Q} \label{eq:SCic_pdc}
\end{align}

The \textbf{reactive power} injection of the IC is described identically to equation \eqref{eq:SC_7b}, where the indices are suitably updated. Therefore, the linear SC model for the reactive power injection in both the $P-Q$ and the $V_{dc} - Q$ operating mode is given by \eqref{eq:SCic_q}.
\begin{align}
    &\frac{\partial Q_{l}^{\phi \ast}}{ \partial x}  =
    \left( \ \ \overbar{F}_{l,l}^{\phi ''} + \overbar{H}_{l}^{\phi ''} \right) \frac{\partial \overbar E_{l}^{\phi'}  }{ \partial x}
    +  {\textstyle\sum\limits_{\substack{n \in \mathcal{N} \setminus \{l\} }}} \overbar{F}_{l,n}^{\phi ''}  \frac{\partial \overbar E_{n}^{\phi'}  }{ \partial x} \nonumber \\
    &\qquad  \ \ + \left( -\overbar{F}_{l,l}^{\phi '}  + \overbar{H}_{l}^{\phi '}\right)  \frac{\partial \overbar E_{l}^{\phi''}  }{ \partial x} 
    -  {\textstyle\sum\limits_{\substack{n \in \mathcal{N} \setminus \{l\} }}} \overbar{F}_{l,n}^{\phi '}  \frac{\partial \overbar E_{n}^{\phi''}  }{ \partial x} \nonumber \\
    & \qquad \qquad \qquad \qquad \qquad \qquad \forall l \in \mathcal{L}_{V_{dc}Q}  \cup \mathcal{L}_{PQ}  \label{eq:SCic_q}
\end{align}

\subsection{Unbalanced loading conditions} \label{subsec:unbalanced}
For unbalanced loading conditions, the linear SC model described by equations \eqref{eq:SC_7a}, \eqref{eq:SC_7b}, \eqref{eq:SC_9}, \eqref{eq:SCdc_5}, \eqref{eq:SCdc_6}, \eqref{eq:SCic_v},  \eqref{eq:SCic_p}, \eqref{eq:SCic_pdc} and \eqref{eq:SCic_q} has to be adapted accordingly.
The linear equations for the AC and DC system are written for each phase individually, and therefore, remains the same. Only the expressions of the ICs model have to be adapted. 
PLLs of each IC typically synchronize with the positive sequence component of the AC grid's voltage. Therefore, the voltage's zero and negative sequence components are zero at the AC terminal of the IC and the ICs will, therefore, only inject positive sequence power.

Using the Fortescue transformation, we decompose the three-phase voltages and currents into their symmetrical components as shown in \eqref{eq:T} \footnote{The symmetrical component transformation requires the assumption that the lines' admittance matrices are circular symmetric.}. $\overbar{\mathbf{E}}^{0+-}$ is the vector containing the zero, positive and negative sequence component of the nodal voltage phasor, $\overbar{\mathbf{T}}$ is the transformation matrix, $\overbar{\mathbf{E}}^{abc} = [ \overbar{E}^{\phi}] \ \forall \ \phi \in \{ a,b,c\}$ and $\alpha = e^{\frac{2}{3} \pi \mi}$. 
\small
    \begin{equation}
    \overbar{\mathbf{E}}^{0+-} = \overbar{\mathbf{T}} \overbar{\mathbf{E}}^{abc} \ \ \ \text{with,} \ \ \
            \overbar{\mathbf{T}} 
          = \frac{1}{3} \left[\begin{IEEEeqnarraybox*}[][c]{,c/c/c,}
            1 & 1 & 1 \\
            1 & \alpha & \ \alpha^2\\
            1 & \ \alpha^2 & \alpha
            \end{IEEEeqnarraybox*}\right] \label{eq:T}
    \end{equation}
\normalsize

For the $\mathbf{V_{dc} - Q}$ operating mode, the PF model is adapted in \eqref{eq:SCic_+} for unbalanced loading conditions. Because the ICs only inject the positive sequence power, the active power balance between the AC and DC part is based on the positive sequence voltage $\overbar{E}^{+}$.
\small
\begin{align}
        &\Re  \Big\{ \overbar{E}_{l}^{+} \sum\nolimits_{n \in \mathcal{N}} \underbar Y_{(l,n)}^{ac} \underbar E_{n}^{+} \Big\}  = 
        E_{k}^{\ast} \Big( Y_{(k,k)}^{dc} E_{k}^{\ast} + {\textstyle\sum\limits_{\substack{m \in \mathcal{M} \setminus \{k\} }}}  Y_{(k,m)}^{dc} E_{m} \Big)  \nonumber \\ 
        &  \qquad \qquad  \qquad \qquad \qquad \qquad  \qquad \qquad \quad \forall (l,k) \in \mathcal{L}_{V_{dc}Q}  \label{eq:SCic_+}
\end{align}
\normalsize
\noindent
If we take the partial derivative of \eqref{eq:SCic_+} to $x$, we obtain a linear expression in $\frac{ \partial \overbar E^+}{ \partial x}$. However, the partial derivative of the positive sequence voltages are not part of the vector of unknowns in \eqref{d_dX}. Therefore, $\frac{ \partial \overbar E^+}{ \partial x}$ in \eqref{eq:SCic_+} has to be transformed again to its phase domain to obtain a linear expression dependent only on the unknown of \eqref{d_dX}.

\noindent
We transform the identities $\overbar{F}$ and $\overbar{H}$ to their sequence domain as shown in \eqref{eq:FH_sym}.
Because of the unbalanced nature, the three phases cannot be treated individually anymore, but the electrical quantities become vectors. Therefore, the identities $\overbar{\mathbf{F}}_{i,n}^{0+-}$ and $\overbar{\mathbf{H}}_{i}^{0+-}$ are 3-by-3 matrices where every row refers to the zero, positive and negative sequence components. Furthermore, we use the notation where $\overbar{\mathbf{F}}_{i,n}^{0}$ is the first row related to the homopolar sequence,  $\overbar{\mathbf{F}}_{i,n}^{+}$ is the second row related to the positive sequence and $\overbar{\mathbf{F}}_{i,n}^{-}$ is the third row related to the negative sequence. The partial derivative of the vector $\overbar{\mathbf{E}}^{abc}$ to $x$ is equal to $\frac{\partial \overbar{\mathbf{E}}^{abc}  }{ \partial x} = \left[ \frac{\partial \overbar{{E}}^{a}  }{ \partial x} , \frac{\partial \overbar{{E}}^{b}  }{ \partial x} , \frac{\partial \overbar{{E}}^{c}  }{ \partial x}     \right]^\intercal$.

\begin{align}
&\overbar{\mathbf{F}}_{i,n}^{0+-} =  \overbar{\mathbf{T}}^{-1} \text{diag} \big( \underbar Y_{i,n}^{ac} \overbar{\mathbf{T}} \overbar{\mathbf{E}}_{i}^{abc} \big) \nonumber \\ 
&\overbar{\mathbf{H}}_{i}^{0+-}   =  \overbar{\mathbf{T}}^{-1} \text{diag} \big( {\textstyle\sum\limits_{n \in \mathcal{N}}} \underbar Y_{i,n}^{ac} \underbar{$\mathbf{T}$} \overbar{\mathbf{E}}_{i}^{abc} \big)  \label{eq:FH_sym}
\end{align}

\noindent
Next, using the above identities, a linear expression of the phase-domain partial derivatives,  $\frac{ \partial \overbar E^{\phi}}{ \partial x}$, is formulated for the ICs operating in $V_{dc} - Q$ mode   \eqref{eq:SCic_v0}. 
\begin{align}
    & \Big( F_{(k,k)} + H_{k} \Big) \frac{\partial E_{k}^{\ast}  }{ \partial x}  = -{\textstyle\sum\limits_{\substack{m \in \mathcal{M} \setminus \{k\} }}}  F_{(k,m)} \frac{\partial  E_{m}  }{ \partial x} \nonumber \\
    & \quad + \Big( \overbar{\mathbf{F}}_{l,l}^{+'} + \ \overbar{\mathbf{H}}_{i}^{+ '} \Big) \frac{\partial \overbar{\mathbf{E}}_{l}^{abc'} }{ \partial x}
    \ +  \Big( {\textstyle\sum\limits_{\substack{n \in \mathcal{N} \setminus \{l\} }}} \overbar{\mathbf{F}}_{l,n}^{+'} \ \Big) \frac{\partial \overbar{\mathbf{E}}_{n}^{abc'} }{ \partial x} \nonumber \\
    & \quad + \Big( \overbar{\mathbf{F}}_{l,l}^{+''}  - \overbar{\mathbf{H}}_{i}^{+''} \Big)  \frac{\partial \overbar{\mathbf{E}}_{l}^{abc''}  }{ \partial x}  +  \ \Big( {\textstyle\sum\limits_{\substack{n \in \mathcal{N} \setminus \{l\} }}} \overbar{\mathbf{F}}_{l,n}^{+''} \Big)  \frac{\partial \overbar{\mathbf{E}}_{n}^{abc''}  }{ \partial x} \nonumber \\
    & \qquad \qquad \qquad  \qquad  \qquad \qquad \qquad  \forall (l,k) \in \mathcal{L}_{V_{dc}Q} 
    \label{eq:SCic_v0} 
\end{align}

\noindent
Furthermore, because the IC only injects the positive sequence powers, the negative and zero sequence powers are zero. As already explained in Section \ref{section:PFmodel}, formulating the linear system of equations using the partial derivative of the zero and negative power injection, results in the trivial expression and thus an undetermined problem. Therefore, the derivation of the SC model starts from expressions \eqref{E0}  and \eqref{E-} that are reformulated into the phase-domain as shown in \eqref{eq:SCic_v0_1}:
\begin{subequations}
\begin{align}
    \overbar{E}_{l}^{0'} &=  \overbar{\mathbf{T}}^{0 '} \overbar{\mathbf{E}}_{l}^{abc'} - \overbar{\mathbf{T}}^{0 ''} \overbar{\mathbf{E}}_{l}^{abc''} \ = 0  \\
    \overbar{E}_{l}^{-'} &=  \overbar{\mathbf{T}}^{- '} \overbar{\mathbf{E}}_{l}^{abc'} - \overbar{\mathbf{T}}^{- ''} \overbar{\mathbf{E}}_{l}^{abc''} = 0, \\
    & \qquad \qquad \qquad  \qquad \qquad \qquad \quad \forall l \in \mathcal{L}_{V_{dc}Q} \nonumber
\end{align} \label{eq:SCic_v0_1}
\end{subequations}

\noindent
where e.g. $\overbar{\mathbf{T}}^{0 '}$ refers to the row corresponding to the zero sequence.
Taking the partial derivative of \eqref{eq:SCic_v0_1} gives the linear expression:
\begin{subequations}
\begin{align}
    &\overbar{\mathbf{T}}^{0 '}  \frac{\partial \overbar{\mathbf{E}}_{l}^{abc'}  }{ \partial x}  - \overbar{\mathbf{T}}^{0 ''}  \frac{\partial \overbar{\mathbf{E}}_{l}^{abc''}  }{ \partial x}  = 0 \\
    &\overbar{\mathbf{T}}^{- '}  \frac{\partial \overbar{\mathbf{E}}_{l}^{abc'}  }{ \partial x}  - \overbar{\mathbf{T}}^{- ''}  \frac{\partial \overbar{\mathbf{E}}_{l}^{abc''}  }{ \partial x}  = 0  \label{eq:SCic_v0_2} \\
    & \qquad \qquad \qquad  \qquad  \qquad \qquad \forall l \in \mathcal{L}_{V_{dc}Q} \cup \mathcal{L}_{PQ}  \nonumber
\end{align} \label{eq:SCic_v0_3}
\end{subequations}

For the $\mathbf{P-Q}$ operating mode, equation \eqref{eq:SCic_p} is adapted for unbalanced conditions similarly. In the case that the IC is only injecting positive sequence power, SC model for the active power injection consists of \eqref{eq:SCic_p_UNB} and the linear equations in \eqref{eq:SCic_v0_3} to describe the zero and negative power injection. In the specific case that the IC control loops also allow intentionally negative power injection, \eqref{eq:SCic_p_UNB} is also formulated for the negative sequence replaces the \eqref{eq:SCic_v0_2}.

\begin{align}
&\frac{\partial P_{l}^{p \ast}}{ \partial x}  =
    \Big( \overbar{\mathbf{F}}_{l,l}^{+ '} + \ \overbar{\mathbf{H}}_{i}^{+ '} \Big) \frac{\partial \overbar{\mathbf{E}}_{l}^{abc'} }{ \partial x}
    \ +  \Big( {\textstyle\sum\limits_{\substack{n \in \mathcal{N} \setminus \{l\} }}} \overbar{\mathbf{F}}_{l,n}^{+ '} \ \Big) \frac{\partial \overbar{\mathbf{E}}_{n}^{abc'} }{ \partial x} \nonumber \\
    & \qquad \ \  + \Big( \overbar{\mathbf{F}}_{l,l}^{+ ''}  - \overbar{\mathbf{H}}_{i}^{+ ''} \Big)  \frac{\partial \overbar{\mathbf{E}}_{l}^{abc''}  }{ \partial x}  +  \Big( {\textstyle\sum\limits_{\substack{n \in \mathcal{N} \setminus \{l\} }}} \overbar{\mathbf{F}}_{l,n}^{+ ''} \Big)  \frac{\partial \overbar{\mathbf{E}}_{n}^{abc''}  }{ \partial x} 
    \nonumber \\
    &  \qquad \qquad \qquad \qquad \qquad \qquad \qquad  \qquad \qquad  \forall l \in \mathcal{L}_{PQ}  \label{eq:SCic_p_UNB}
\end{align}

The reactive power injections of both operation modes of the IC under unbalanced conditions is formulated in \eqref{eq:SCic_q_UNB}.
\begin{subequations}
\begin{align}
    &\frac{\partial Q_{l}^{\phi \ast}}{ \partial x}  =
    \Big( \overbar{\mathbf{F}}_{l,l}^{+''} + \ \overbar{\mathbf{H}}_{i}^{+ ''} \Big) \frac{\partial \overbar{\mathbf{E}}_{l}^{abc'} }{ \partial x}
    \ +  \Big( {\textstyle\sum\limits_{\substack{n \in \mathcal{N} \setminus \{l\} }}} \overbar{\mathbf{F}}_{l,n}^{+ ''} \ \Big) \frac{\partial \overbar{\mathbf{E}}_{n}^{abc'} }{ \partial x} \nonumber \\
    & \qquad \ \ + \Big( \overbar{\mathbf{F}}_{l,l}^{+ '}  - \overbar{\mathbf{H}}_{i}^{+ '} \Big)  \frac{\partial \overbar{\mathbf{E}}_{l}^{abc''}  }{ \partial x}  +  \Big( {\textstyle\sum\limits_{\substack{n \in \mathcal{N} \setminus \{l\} }}} \overbar{\mathbf{F}}_{l,n}^{+ '} \Big)  \frac{\partial \overbar{\mathbf{E}}_{n}^{abc''}  }{ \partial x} 
     \\
    &\overbar{\mathbf{T}}^{0 ''}  \frac{\partial \overbar{\mathbf{E}}_{l}^{abc'}  }{ \partial x}  + \overbar{\mathbf{T}}^{0 '}  \frac{\partial \overbar{\mathbf{E}}_{l}^{abc''}  }{ \partial x}  = 0  \\
    &\overbar{\mathbf{T}}^{- ''}  \frac{\partial \overbar{\mathbf{E}}_{l}^{abc'}  }{ \partial x}  + \overbar{\mathbf{T}}^{- '}  \frac{\partial \overbar{\mathbf{E}}_{l}^{abc''}  }{ \partial x}  = 0  \\
    &\qquad \qquad  \qquad \qquad  \qquad \qquad \qquad  \forall l \in \mathcal{L}_{V_{dc}Q}  \cup \mathcal{L}_{PQ}  \nonumber
\end{align} \label{eq:SCic_q_UNB}
\end{subequations}

\subsection{Linear system of equations}

The closed-form SC model for hybrid AC/DC networks is formulated as equations \eqref{eq:SC_7a}, \eqref{eq:SC_7b}, \eqref{eq:SC_9}, \eqref{eq:SCdc_5}, \eqref{eq:SCdc_6}, \eqref{eq:SCic_v},  \eqref{eq:SCic_pdc}, \eqref{eq:SCic_v0}, \eqref{eq:SCic_v0_3}, \eqref{eq:SCic_p_UNB} and \eqref{eq:SCic_q_UNB}. Regrouping these equations results in a system of equations that is linear with respect to the partial derivatives of the voltage \eqref{eq:X}. The modal is formulated as $\mathbf{A} \  \mathbf{u}(x) = \mathbf{b}(x)$.

\begin{figure}[H]
 \centering
   \includegraphics[width=1\linewidth]{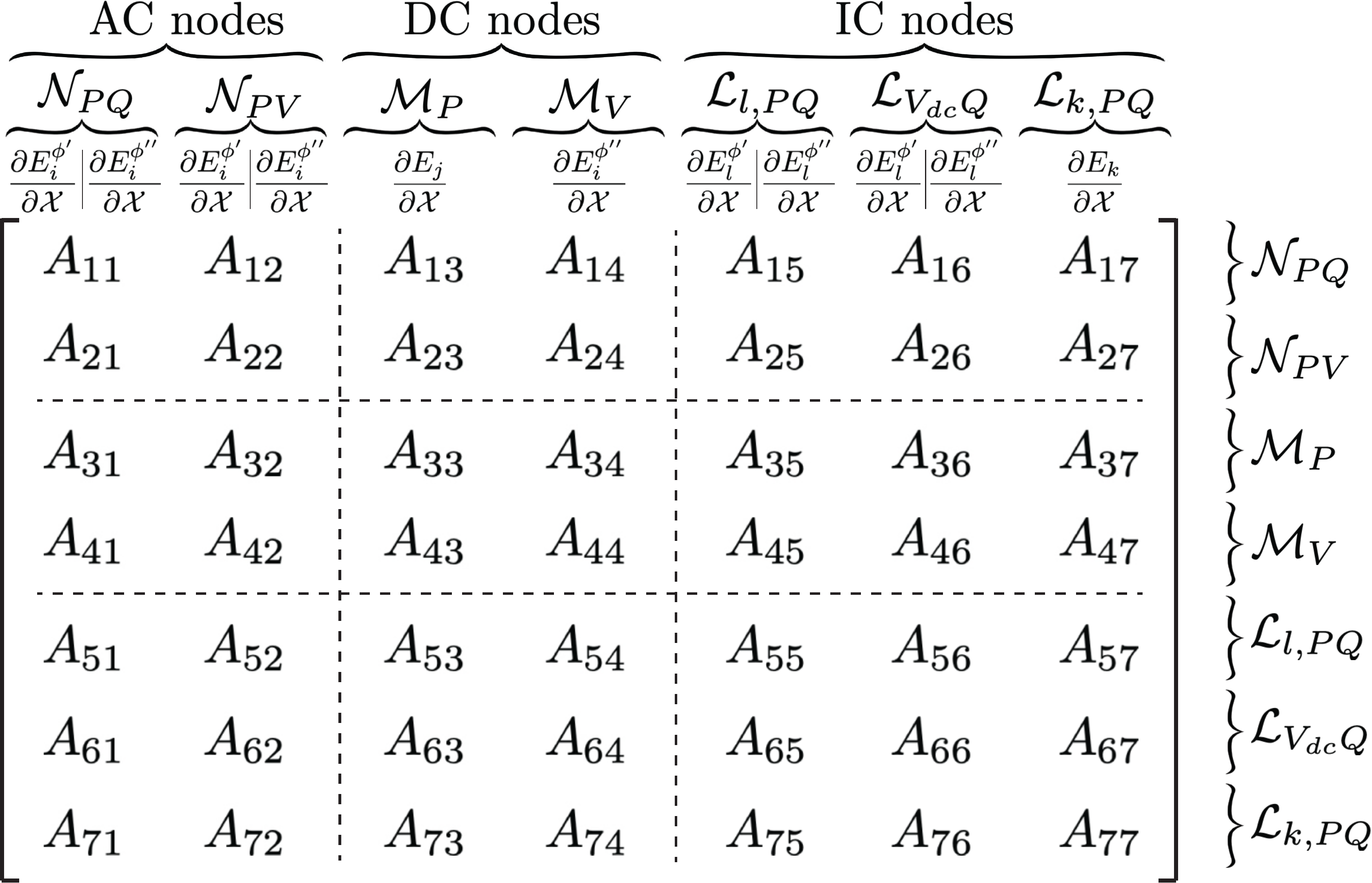}
   \vspace{-0.9cm}
   \begin{align}
     \label{eq:bigA}
 \end{align}
 \end{figure}

The matrix $\mathbf{A}$ is defined in \eqref{eq:bigA} as a $7 \ \times \ 7$ block matrix, one for each type of node in the AC grid, DC grid and IC. The rows represent the linearized PF equations derived above. The columns represent the individual terms of each linearized PF equation, grouped per node type. 
The matrix $\mathbf{A}$ only depends on the state of the grid i.e., the nodal voltage at every node, and the admittance matrix. Furthermore, is it independent of the control variables $\mathcal{X}$ and thus has to be computed only once. We do not give explicitly the expressions for $A_{ij}$ since they are trivially derived from \eqref{eq:SC_7a}, \eqref{eq:SC_7b}, \eqref{eq:SC_9}, \eqref{eq:SCdc_5}, \eqref{eq:SCdc_6}, \eqref{eq:SCic_v},  \eqref{eq:SCic_pdc}, \eqref{eq:SCic_v0}, \eqref{eq:SCic_v0_3}, \eqref{eq:SCic_p_UNB} and \eqref{eq:SCic_q_UNB}.

The vector $\mathbf{b}$ depends on $x$ is the left-hand side of the linearized equations evaluated for $x \in \mathcal{X}$ 
and will thus have to be recomputed computed for every control variable. The vector is computed following to the rules defined in \eqref{eq:rules_1} and \eqref{eq:rules_2}. The method and the construction of matrix $\mathbf{A}$ and $\mathbf{u}$ is publicly available on \url{https://github.com/DESL-EPFL}.

\vspace{0.3cm}
\begin{theorem}
The analytical model for the computation of the SCs for hybrid AC/DC grids, given by equations \eqref{eq:SC_7a}, \eqref{eq:SC_7b}, \eqref{eq:SC_9}, \eqref{eq:SCdc_5}, \eqref{eq:SCdc_6}, \eqref{eq:SCic_v},  \eqref{eq:SCic_pdc}, \eqref{eq:SCic_v0}, \eqref{eq:SCic_v0_3}, \eqref{eq:SCic_p_UNB} and \eqref{eq:SCic_q_UNB} has a unique solution of any operating point where the PF Jacobian of the hybrid AC/DC network is invertible. This follows directly from the theorem in \cite{paolone2015optimal} that proves the uniqueness of the solution of the SCs for a solely AC network. The proof of the corollary is given in Appendix \ref{appendix:A}.
\end{theorem}


\subsection{Current sensitivity coefficients}

The branch current sensitivity coefficients can be obtained by using the network admittance matrix. Assuming that the branches can be represented as a $\pi$-model equivalent, the AC current flow between the buses $i$ and $n$ and the DC current flow between buses $j$ and $m$ is expressed as:
\begin{align}
    &\overbar{I}_{i,n} = \overbar{Y}_{i,n}^{ac} \left( \overbar{E}_{i} - \overbar{E}_{n} \right) + \overbar{Y}_{i_{0}}^{ac} \overbar{E}_{i}, \nonumber \\
    &I_{j,m} = Y_{j,m}^{dc} \left( E_{j} - E_{m} \right),
\end{align}
\noindent
where the complex voltage SC are computed using  \eqref{SCidentity2} and $\overbar{Y}_{i_{0}}^{ac}$ is the shunt element of the branch $\pi$-model model in node $i$. Therefore, the branch current sensitivity coefficients are described as \eqref{eq:SC_I}.
\begin{align}
     \frac{\partial  {\overbar{I}_{i,n}} }{ \partial x} = & \overbar{Y}_{i,n}^{ac} (\frac{\partial  {\overbar{E}_{i}} }{ \partial x} - \frac{\partial  {\overbar{E}_{n}} }{ \partial x}) + \overbar{Y}_{i_{0}}^{ac} \frac{\partial  {\overbar{E}_{i}} }{ \partial x}  \\
     \frac{\partial {I_{j,m}}}{ \partial x} = & Y_{j,m}^{dc} (\frac{\partial {E_{j}}}{ \partial x} - \frac{\partial {E_{m}}}{ \partial x}) \label{eq:SC_I}
\end{align}

\section{Numerical validation}
\label{section:Num_validation}

The numerical validation of the closed-form expression of the sensitivity coefficients is performed in accurate time-domain simulation in the EMTP-RV environment \cite{emtp}. The SCs are numerically computed by applying a small perturbation to the network and observing the response. The perturbation is realized by slightly changing the setpoint of each controllable variable, once at a time. This is repeated for every node type to compare the numerically computed SCs with the analytical ones. The voltage SCs of the AC and DC nodes are computed as in \eqref{eq:vali}, where $t=t_0$ is before the perturbation and $t=t_{1}$ is after when the system reached its post-perturbation steady-state\footnote{The value of $t$ is case-dependent and assessed by the user as a function of the perturbation type and its location.}. These numerical SCs are considered as the ground truth for the validation shown later in this section. 

\begin{subequations}
\begin{align}
    &\frac{\Delta \overbar{E}_i^{'}}{\Delta x} = \frac{\overbar{E}_i^{t_0} - \overbar{E}_i^{t_{1}}}{x^{t_0} - x^{t_{1}}} \qquad \forall x \in \mathcal{X}, i \in \mathcal{N} \\
    &\frac{\Delta {E}_j^{'}}{\Delta x} = \frac{{E}_j^{t_0} - {E}_i^{t_{1}}}{x^{t_0} - x^{t_{1}}} \qquad \forall x \in \mathcal{X}, j \in \mathcal{M}
\end{align}  \label{eq:vali}
\end{subequations}

\subsection{Simulation Setup}
\begin{figure}[H]
 \centering
   \includegraphics[width=0.9\linewidth]{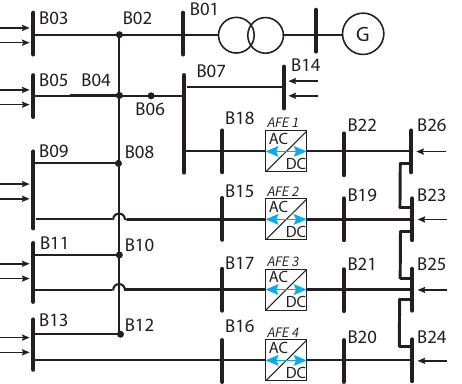}
   \caption{Topology of the hybrid AC/DC microgrid used for the numerical validation of the closed-form computation of the SC.}
   \label{fig:Mgrid}
 \end{figure}

The topology of the hybrid AC/DC grid used for the numerical simulation is shown in Figure \ref{fig:Mgrid}. The grid consists of the low voltage CIGRE benchmark grid \cite{Barsali2005CIGRETF} that is suitably extended with a DC grid. The AC network consists of 18 nodes and has a base voltage of \SI{400}{\volt}. The network is connected to the medium voltage grid in bus $B01$. The DC grid consists of 8 nodes and has a base voltage of \SI{800}{\volt}. The base power for both networks is \SI{100}{\kilo \watt}. The two grid are interconnected using 4 ICs: \textit{ IC 1} and \textit{IC 4} operate under voltage control mode ($V_{dc}-Q_{ac}$) and \textit{IC 2 } and \textit{IC 3} operate under power control mode ($P_{ac}-Q_{ac}$). 
The boundary conditions of the simulation, i.e. the power injections in the P(Q) nodes and the voltage profile in (P)V and IC nodes, are sampled from real measurements of the hybrid AC/DC microgrid that is available at the Distributed Electrical System Laboratory (DESL) at the EPFL (see \cite{willem_SC_exp} for further details). This hybrid microgrid has the exact same topology and parameters as the one used in the simulation. Furthermore, the unbalanced loading conditions in the AC network are created by a difference of \num{0.2} p.u. between the phases of bus $B11$. The grid's parameters and resources are presented in \cite{willem_SE_exp} The simulation model is made publicly available on the DESL GitHub page\footnote{\url{https://github.com/DESL-EPFL}} \cite{github}.

\subsection{Numerical validation}
\label{section:vali}

The results of the numerical validation are summarised and presented in Table \ref{tab:results}. The mean and maximum error between the ground truth, obtained in the EMTP-RV simulation, and the analytically computed voltage SCs are shown for the different node types in hybrid AC/DC networks. The ground truth is obtained by perturbing the setpoints of the controllable resources (i.e. $\Delta \SI{500}{W} , \ \Delta Q = \SI{500}{VAr} $ and $\Delta V = \SI{4}{V}  $) and analysing the results as shown in \eqref{eq:vali}. We can see the that analytical solution is very accurate with a mean and max error are resp. in the order of \num{1e-4} p.u. and \num{1e-3} p.u. Therefore, we can assert that the analytical method for the computation of the voltage SCs in hybrid AC/DC networks provides reliable estimates. Furthermore, the closed-form computation of the voltage SCs takes on average \SI{50}{\milli \second} for the three-phase 26-node hybrid grid presented in \ref{fig:Mgrid}.

\begin{table}[H] \centering 
\begin{tabular}{llllll}
\hline
Bus     & Network          &  Bus type           &   & Mean error [p.u.]   & Max error [p.u.]\\ \hline
$P_9$    &\textit{AC}      & $P - Q$             &   & -2.70e-4    &  \ 1.46e-3       \\
$Q_9$    &\textit{AC}      & $P - Q$             &   & -2.51e-5   &  -3.61e-3         \\
$Q_{18}$ &\textit{IC}      & $V_{dc} - Q$        &   & \ 1.60e-4   &  \ 2.40e-3        \\
$E_{22}$ &\textit{IC}      & $V_{dc} - Q$        &   &  -1.03e-4   &  \ 2.90e-3        \\
$P_{17}$ &\textit{IC}      & $P - Q$             &   &  \ 4.45e-5  &  -1.45e-3        \\
$P_{23}$ &\textit{DC}      & $P$                 &   &  -5.57e-5   &  -8.91e-4        \\ \hline  
\end{tabular} 
\caption{The mean and maximum error between the EMTP-RV simulation and the analytical solution for the different node types}
\label{tab:results}
\end{table}

\section{Conclusion}
\label{section:Conclusion}

In this paper, we present a method for the closed-form analytical computation of the power flow SCs in unbalanced hybrid AC/DC networks. The SCs, defined as the partial derivaties of the nodal voltage with respect the the power injections, allow to formulate the grid constraints in the OPF problem in a fully linear way. Therefore, leveraging the real-time control of hybrid AC/DC networks. The proposed method is inspired by an existing SC computation process of solely AC networks and extended for hybrid networks using a unified PF model that accounts for the AC grid, DC grid and the various operation modes of the ICs. The uniqueness of the proposed SCs computational model is shown in a formal proof.
Furthermore, the model is numerically validated in an accurate time-domain simulation in the EMTP-RV environment. It is shown that the accuracy of the SCs is in the order of \num{1e-4} p.u. and has a computation time of around $\SI{50}{\milli \second}$ for the considered 26-node hybrid network.

\appendices
\section{Proof of uniqueness}
\label{appendix:A}

The system of equations 
\eqref{eq:SC_7a}, \eqref{eq:SC_7b}, \eqref{eq:SC_9}, \eqref{eq:SCdc_5}, \eqref{eq:SCdc_6}, \eqref{eq:SCic_v},  \eqref{eq:SCic_pdc}, \eqref{eq:SCic_v0}, \eqref{eq:SCic_v0_3}, \eqref{eq:SCic_p_UNB} and \eqref{eq:SCic_q_UNB}
is linear with respect to the partial derivatives of the real and imaginary part of the nodal voltage. Therefore, the uniqueness of the solution, i.e. the voltage SCs, can be proved by showing that the homogeneous system of equations only has the trivial solution. The corollary is based on the main theorem in \cite{paolone2015optimal}. Because of space limitations, the proof is only shown for the system under balanced loading conditions.

Similarly to what has been done in \cite{paolone2015optimal}, The linear system of equations given in Section \ref{section:SCmodel} can be written as in \eqref{eq:proof_homo} where $\overbar{\Delta}^{\prime}$ and $\overbar{\Delta}^{\prime \prime}$ are the unknown real and imaginary partial derivatives of the nodal voltages (in this case for both the AC and DC grids). 

\small
\allowdisplaybreaks
\begin{subequations} \label{eq:proof_homo}
\begin{align} 
    &0 =
    \overbar{H}_{i}^{\phi '}  \overbar{\Delta}^{\prime}_{i}
     +  {\textstyle\sum\limits_{n \in \mathcal{N} }} \overbar{F}_{i,n}^{\phi '}  \overbar{\Delta}^{\prime}_{n}  - \overbar{H}_{i}^{\phi ''} \overbar{\Delta}^{\prime \prime}_{i}
    +  {\textstyle\sum\limits_{n \in \mathcal{N} }} \overbar{F}_{i,n}^{\phi ''}  \overbar{\Delta}^{\prime \prime}_{n} \nonumber\\
    & \qquad  \qquad  \qquad  \qquad  \qquad  \qquad  \qquad  \qquad \forall i \in \mathcal{N}_{PQ} \cup \mathcal{N}_{PV} \\
    &0  =
    \overbar{H}_{i}^{\phi ''} \overbar{\Delta}^{\prime}_{i}
    +  {\textstyle\sum\limits_{n \in \mathcal{N} }} \overbar{F}_{i,n}^{\phi '}  \overbar{\Delta}^{\prime}_{n}   + \overbar{H}_{i}^{\phi '}  \overbar{\Delta}^{\prime \prime }_{i}
    -  {\textstyle\sum\limits_{n \in \mathcal{N} }} \overbar{F}_{i,n}^{\phi ''}  \overbar{\Delta}^{\prime \prime}_{n} \nonumber\\
    &  \qquad  \qquad  \qquad  \qquad  \qquad  \qquad  \qquad  \qquad\forall i \in \mathcal{N}_{PQ} \\
    & 0 = \overbar{E}_{i}^{\phi '} \overbar{\Delta}^{\prime}_{i} +  \overbar{E}_{i}^{\phi ''}\overbar{\Delta}^{\prime \prime}_{i},    \qquad  \qquad  \qquad \hspace{2pt} \forall i \in \mathcal{N}_{PV} \\
    &0  =
     {H}_{j}  \overbar{\Delta}_{j}
    \ +  {\textstyle\sum\limits_{m \in \mathcal{M} }} {F}_{j,m}  {\Delta}_{m}
      \qquad  \qquad  \forall j \in \mathcal{M}_{P} \\
    &0  =  {\Delta}_{j},
      \qquad  \qquad  \qquad  \qquad  \qquad  \qquad \hspace{3pt} \forall j \in \mathcal{M}_{V} \\
    & 0 =
    {H}_{k} {\Delta}_{k} +  {\textstyle\sum\limits_{m \in \mathcal{M} }} {F}_{k,m}  {\Delta}_{m} 
     \qquad  \qquad   \hspace{1pt} \forall k \in \mathcal{L}_{PQ} \\
    & 0 =
       \overbar{{H}}_{i}^{\phi '}  {\overbar{\Delta}}^{'}_{l}
     +  {\textstyle\sum\limits_{n \in \mathcal{N} }} \overbar{{F}}_{l,n}^{\phi '} {\overbar{\Delta}}^{'}_{n}
     - \overbar{{H}}_{i}^{ \phi ''}  {\overbar{\Delta}}^{''}_{l}  
     +  {\textstyle\sum\limits_{n \in \mathcal{N} }} \overbar{{F}}_{l,n}^{\phi ''}  {\overbar{\Delta}}^{''}_{n} \nonumber \\
     &  \qquad  \qquad  \qquad  \qquad  \qquad  \qquad  \qquad  \qquad  \forall l \in \mathcal{L}_{PQ}, \\
    & 0 =
      \overbar{{H}}_{i}^{\phi ''} {\overbar{\Delta}}^{'}_{l} +  {\textstyle\sum\limits_{n \in \mathcal{N} }} \overbar{{F}}_{l,n}^{\phi ''} {\overbar{\Delta}}^{'}_{n}
      + \overbar{{H}}_{i}^{\phi '}   {\overbar{\Delta}}^{''}_{l}  -   {\textstyle\sum\limits_{n \in \mathcal{N} }} \overbar{{F}}_{l,n}^{\phi '}  {\overbar{\Delta}}^{''}_{n}  \nonumber\\
    &  \qquad  \qquad  \qquad  \qquad  \qquad  \qquad  \qquad  \qquad \forall l \in \mathcal{L}_{E_{dc}Q}  \cup \mathcal{L}_{PQ} \\
    & 0  = 
     \overbar{{H}}_{i}^{\phi '} {\overbar{\Delta}}^{'}_{l}
     +  {\textstyle\sum\limits_{n \in \mathcal{N} }} \overbar{{F}}_{l,n}^{\phi '}  {\overbar{\Delta}}^{'}_{n}   - \overbar{{H}}_{i}^{\phi ''}  {\overbar{\Delta}}^{''}_{l} +  {\textstyle\sum\limits_{n \in \mathcal{N} }} \overbar{{F}}_{l,n}^{\phi ''}  {\overbar{\Delta}}^{''}_{n} \nonumber \\
     &- {\textstyle\sum\limits_{m \in \mathcal{M} \setminus \{k\} }}  F_{k,m} {\Delta}_{m} 
     ,  \qquad  \qquad  \qquad  \ \hspace{2pt}   \forall (l,k) \in \mathcal{L}_{E_{dc}Q} 
\end{align}
\end{subequations}
\normalsize
\noindent
We want to show that the only solution to this system is the trivial one i.e., $\overbar{\Delta}^{\prime} = \overbar{\Delta}^{\prime \prime } = 0$. Let's consider two hybrid AC/DC networks with the same topology and the same parameters i.e., with identical $\overbar{\mathbf{Y}}^{ac}$ and $\mathbf{Y}^{dc}$. The voltages in \textit{network I} are given in \eqref{eq:proof_V1} and the voltages in  \textit{network II} are given in \eqref{eq:proof_V2} where $\epsilon$ is a positive real number.
\small
\begin{subequations} \label{eq:proof_V1}
\begin{align} 
    &\overbar{E}^{I}_{i} = \overbar{E}_{i} + \epsilon \overbar{\Delta}_{i}, \qquad  &&\forall i \in \mathcal{N}_{PQ} \cup \mathcal{N}_{PV} \qquad \\
    &{E}^{I}_{j} = {E}_{j} + \epsilon {\Delta}_{j}, &&\forall j \in \mathcal{M}_{P} \\
    &\overbar{E}^{1}_{l} = \overbar{E}_{l} + \epsilon \overbar{\Delta}_{l}, &&\forall l \in \mathcal{L}_{PQ} \cup \mathcal{L}_{V_{dc}Q} \\
    &{E}^{I}_{k} = {E}_{k} + \epsilon {\Delta}_{k}, &&\forall k \in \mathcal{L}_{PQ} \\
    &{E}^{I}_{k} = {E}_{k},  &&\forall k \in \mathcal{L}_{V_{dc}Q} \\
    &{E}^{I}_{j} = {E}_{j},  &&\forall j \in \mathcal{M}_{V}
\end{align}
\end{subequations}
\begin{subequations} \label{eq:proof_V2}
\begin{align} 
    &\overbar{E}^{II}_{i} = \overbar{E}_{i} - \epsilon \overbar{\Delta}_{i}, \qquad  &&\forall i \in \mathcal{N}_{PQ} \cup \mathcal{N}_{PV} \qquad \\
    &{E}^{II}_{j} = {E}_{j} - \epsilon {\Delta}_{j}, &&\forall j \in \mathcal{M}_{P} \\
    &\overbar{E}^{II}_{l} = \overbar{E}_{l} - \epsilon \overbar{\Delta}_{l}, &&\forall l \in \mathcal{L}_{PQ} \cup \mathcal{L}_{V_{dc}Q} \\
    &{E}^{II}_{k} = {E}_{k} -  \epsilon {\Delta}_{k}, &&\forall k \in \mathcal{L}_{PQ} \\
    &{E}^{II}_{k} = {E}_{k},  &&\forall k \in \mathcal{L}_{V_{dc}Q} \\
    &{E}^{II}_{j} = {E}_{j},  &&\forall j \in \mathcal{M}_{V}
\end{align}
\end{subequations}
\normalsize
\noindent
Using \eqref{eq:PFmodel} we can formulate the PF equations for the two networks. We use the following notation for the complex AC voltages: $\overbar{E}^{I}_{i} = \overbar{E}^{I'}_{i} + j \overbar{E}^{I''}_{i}$ and the identity $F$, defined in \eqref{eq:SC_6},  is reformulated as  $\overbar{F}^{I,\phi}_{i,n} = \overbar{F}^\phi_{i,n} + \epsilon \overbar{F}^{\Delta,\phi}_{i,n}$, where $\overbar{F}^{\Delta}_{i,n} = \Delta_i \underbar Y_{i,n}^{ac}$. The PF equations for the first network are given in \eqref{eq:proof_PF1} and for the second network in \eqref{eq:proof_PF2}. The equations are constructed starting from \eqref{eq:PFmodel}, the voltages \eqref{eq:proof_V1} and \eqref{eq:proof_V2} are substituted, and the equations are split into their real and imaginary part.

\begin{figure*}
    
\small
\allowdisplaybreaks
\begin{subequations}\label{eq:proof_PF1}
\begin{align} 
      &P^{I,\phi}_{i} =
    \overbar{E}_i^{\phi '} \overbar{H}_{i}^{\phi '} +               \epsilon   {\textstyle\sum\limits_{n \in \mathcal{N} }} \overbar{F}_{i,n}^{\phi '}          \underbar{$\Delta$}_{n}^{'} 
    +\epsilon \overbar{\Delta}_i^{\phi '} \overbar{H}_{i}^{\phi '} +\epsilon^2 {\textstyle\sum\limits_{n \in \mathcal{N} }} \overbar{F}_{i,n}^{\Delta \phi '}   \underbar{$\Delta$}_{n}^{'}
    -\overbar{E}_i^{\phi ''} \overbar{H}_{i}^{\phi ''} -            \epsilon   {\textstyle\sum\limits_{n \in \mathcal{N} }} \overbar{F}_{i,n}^{\phi ''}         \underbar{$\Delta$}_{n}^{''}
    -\epsilon \overbar{E}_i^{\phi ''} \overbar{H}_{i}^{\phi ''} -   \epsilon^2 {\textstyle\sum\limits_{n \in \mathcal{N} }} \overbar{F}_{i,n}^{\Delta \phi ''}  \underbar{$\Delta$}_{n}^{''}
\nonumber \\
    &\qquad \qquad \qquad \qquad \qquad \qquad \qquad \qquad \qquad \qquad \qquad \qquad \qquad \qquad \qquad \qquad \qquad \qquad \qquad \qquad \qquad \quad \ \hspace{2pt} \forall i \in \mathcal{N}_{PQ} \cup \mathcal{N}_{PV} 
    \\
    &Q^{I,\phi}_{i} =
    \overbar{E}_i^{\phi '} \overbar{H}_{i}^{\phi ''} +               \epsilon   {\textstyle\sum\limits_{n \in \mathcal{N} }} \overbar{F}_{i,n}^{\phi '} \underbar{$\Delta$}_{n}^{''} +
    \epsilon \overbar{\Delta}_i^{\phi '} \overbar{H}_{i}^{\phi ''} + \epsilon^2 {\textstyle\sum\limits_{n \in \mathcal{N} }} \overbar{F}_{i,n}^{\Delta \phi '} \underbar{$\Delta$}_{n}^{''} +
    \overbar{E}_i^{\phi ''} \overbar{H}_{i}^{\phi '} +            \epsilon   {\textstyle\sum\limits_{n \in \mathcal{N} }} \overbar{F}_{i,n}^{\phi ''} \underbar{$\Delta$}_{n}^{'} +
    \epsilon \overbar{E}_i^{\phi ''} \overbar{H}_{i}^{\phi '} +   \epsilon^2 {\textstyle\sum\limits_{n \in \mathcal{N} }} \overbar{F}_{i,n}^{\Delta \phi ''} \underbar{$\Delta$}_{n}^{'}
\nonumber \\
    &\qquad \qquad \qquad \qquad \qquad \qquad \qquad \qquad \qquad \qquad \qquad \qquad \qquad \qquad \qquad \qquad \qquad \qquad \qquad \qquad \qquad \quad \ \hspace{2pt}  \forall i \in \mathcal{N}_{PQ} 
    \\
    & \lvert \overbar{E}_{i}^{I \phi} \rvert ^2 = (\overbar{E}_{i}^{\phi '})^2 + 2 \epsilon \overbar{E}_{i}^{\phi '} \overbar{\Delta}^{'}_{i} +  \epsilon^2 ( \overbar{\Delta}^{'}_{i})^2 + (\overbar{E}_{i}^{\phi ''})^2 + 2 \epsilon \overbar{E}_{i}^{\phi ''} \overbar{\Delta}^{''}_{i} +  \epsilon^2 ( \overbar{\Delta}^{''}_{i})^2
\qquad \qquad \qquad \qquad \qquad \qquad \hspace{0pt}  \forall i \in \mathcal{N}_{PV} 
    \\
    &P^{I}_{j} =
    {E}_j {H}_{j} +               \epsilon   {\textstyle\sum\limits_{m \in \mathcal{M} }} {F}_{j,m} {\Delta}_{m} +
    \epsilon {\Delta}_j {H}_{j} + \epsilon^2 {\textstyle\sum\limits_{m \in \mathcal{M} }} {F}_{j,m}^{\Delta} {\Delta}_{m}
\qquad \qquad \qquad \qquad \qquad \qquad \qquad \qquad \qquad \hspace{6pt} \forall j \in \mathcal{M}_{P}, 
    \\
    &P^{I}_{k} =
    {E}_k {H}_{k} +               \epsilon   {\textstyle\sum\limits_{m \in \mathcal{M} }} {F}_{k,m} {\Delta}_{m} +
    \epsilon {\Delta}_k {H}_{k} + \epsilon^2 {\textstyle\sum\limits_{m \in \mathcal{M} }} {F}_{k,m}^{\Delta} {\Delta}_{m}
\qquad \qquad \qquad \qquad \qquad \qquad \qquad \qquad \qquad \hspace{1pt} \forall k \in \mathcal{L}_{E_{dc}Q} 
    \\
    &P^{I,\phi}_{l} =
    \overbar{E}_l{\phi '} \overbar{H}_{l}^{\phi '} +                \epsilon   {\textstyle\sum\limits_{n \in \mathcal{N} }} \overbar{F}_{l,n}^{\phi '} \underbar{$\Delta$}_{n}^{'} +
    \epsilon \overbar{\Delta}_l^{\phi '} \overbar{H}_{l}^{\phi '} + \epsilon^2 {\textstyle\sum\limits_{n \in \mathcal{N} }} \overbar{F}_{l,n}^{\Delta \phi '} \underbar{$\Delta$}_{n}^{'}
    -\overbar{E}_l^{\phi ''} \overbar{H}_{l}^{\phi ''} -            \epsilon   {\textstyle\sum\limits_{n \in \mathcal{N} }} \overbar{F}_{l,n}^{\phi ''} \underbar{$\Delta$}_{n}^{''}
    -\epsilon \overbar{E}_l^{\phi ''} \overbar{H}_{l}^{\phi ''} -   \epsilon^2 {\textstyle\sum\limits_{n \in \mathcal{N} }} \overbar{F}_{l,n}^{\Delta \phi ''} \underbar{$\Delta$}_{n}^{''}
\nonumber \\
    &\qquad \qquad \qquad \qquad \qquad \qquad   \forall l \in \mathcal{L}_{PQ}, 
    \\
    &Q^{I,\phi}_{l} =
    \overbar{E}_l^{\phi '} \overbar{H}_{l}^{\phi ''} +               \epsilon   {\textstyle\sum\limits_{n \in \mathcal{N} }} \overbar{F}_{l,n}^{\phi '} \underbar{$\Delta$}_{n}^{''} +
    \epsilon \overbar{\Delta}_l^{\phi '} \overbar{H}_{l}^{\phi ''} + \epsilon^2 {\textstyle\sum\limits_{n \in \mathcal{N} }} \overbar{F}_{l,n}^{\Delta \phi '} \underbar{$\Delta$}_{n}^{''} +
    \overbar{E}_l^{\phi ''} \overbar{H}_{l}^{\phi '} +               \epsilon   {\textstyle\sum\limits_{n \in \mathcal{N} }} \overbar{F}_{l,n}^{\phi ''} \underbar{$\Delta$}_{n}^{'} +
    \epsilon \overbar{E}_l^{\phi ''} \overbar{H}_{l}^{\phi '} +      \epsilon^2 {\textstyle\sum\limits_{n \in \mathcal{N} }} \overbar{F}_{l,n}^{\Delta \phi ''} \underbar{$\Delta$}_{n}^{'}
\nonumber \\
    &\qquad \qquad \qquad \qquad \qquad \qquad \qquad \qquad \qquad \qquad \qquad \qquad \qquad \qquad \qquad \qquad \qquad \qquad \qquad \qquad \qquad \quad \ \hspace{2pt} \forall l \in \mathcal{L}_{E_{dc}Q}  \cup \mathcal{L}_{PQ} 
    \\
    & E_{k}^{I \ast} Y_{k,k}^{dc} E_{k}^{I \ast} +   {\textstyle\sum\limits_{\substack{m \in \mathcal{M} \setminus \{k\} }}} Y_{k,m}^{dc} E_{m} + \epsilon {\textstyle\sum\limits_{\substack{m \in \mathcal{M} \setminus \{k\} }}} Y_{k,m}^{dc} \Delta_{m} 
    = \qquad \qquad \qquad \qquad \qquad \qquad \qquad \qquad \qquad \hspace{7pt}
    \forall (l,k) \in \mathcal{L}_{E_{dc}Q} \nonumber \\
    & \qquad \overbar{E}_l{\phi '} \overbar{H}_{l}^{\phi '} +                \epsilon   {\textstyle\sum\limits_{n \in \mathcal{N} }} \overbar{F}_{l,n}^{\phi '} \underbar{$\Delta$}_{n}^{'} +
    \epsilon \overbar{\Delta}_l^{\phi '} \overbar{H}_{l}^{\phi '} + \epsilon^2 {\textstyle\sum\limits_{n \in \mathcal{N} }} \overbar{F}_{l,n}^{\Delta \phi '} \underbar{$\Delta$}_{n}^{'}
    -\overbar{E}_l^{\phi ''} \overbar{H}_{l}^{\phi ''} -            \epsilon   {\textstyle\sum\limits_{n \in \mathcal{N} }} \overbar{F}_{l,n}^{\phi ''} \underbar{$\Delta$}_{n}^{''}
    -\epsilon \overbar{E}_l^{\phi ''} \overbar{H}_{l}^{\phi ''} -   \epsilon^2 {\textstyle\sum\limits_{n \in \mathcal{N} }} \overbar{F}_{l,n}^{\Delta \phi ''} \underbar{$\Delta$}_{n}^{''} \label{eq:proof_PF1_vdc}
\end{align}
\end{subequations}
\normalsize
\end{figure*}

\begin{figure*}
\small
\begin{subequations}\label{eq:proof_PF2}
\begin{align} 
      &P^{II,\phi}_{i} =
    \overbar{E}_i^{\phi '} \overbar{H}_{i}^{\phi '} -               \epsilon   {\textstyle\sum\limits_{n \in \mathcal{N} }} \overbar{F}_{i,n}^{\phi '}          \underbar{$\Delta$}_{n}^{'} 
    -\epsilon \overbar{\Delta}_i^{\phi '} \overbar{H}_{i}^{\phi '} +\epsilon^2 {\textstyle\sum\limits_{n \in \mathcal{N} }} \overbar{F}_{i,n}^{\Delta \phi '}   \underbar{$\Delta$}_{n}^{'}
    -\overbar{E}_i^{\phi ''} \overbar{H}_{i}^{\phi ''} +            \epsilon   {\textstyle\sum\limits_{n \in \mathcal{N} }} \overbar{F}_{i,n}^{\phi ''}         \underbar{$\Delta$}_{n}^{''}
    +\epsilon \overbar{E}_i^{\phi ''} \overbar{H}_{i}^{\phi ''} -   \epsilon^2 {\textstyle\sum\limits_{n \in \mathcal{N} }} \overbar{F}_{i,n}^{\Delta \phi ''}  \underbar{$\Delta$}_{n}^{''} \nonumber \\
    &\qquad \qquad \qquad \qquad \qquad \qquad \qquad \qquad \qquad \qquad \qquad \qquad \qquad \qquad \qquad \qquad \qquad \qquad \qquad \qquad \qquad \quad \ \hspace{2pt} \forall i \in \mathcal{N}_{PQ} \cup \mathcal{N}_{PV} 
    \\
    &Q^{II,\phi}_{i} =
    \overbar{E}_i^{\phi '} \overbar{H}_{i}^{\phi ''} -               \epsilon   {\textstyle\sum\limits_{n \in \mathcal{N} }} \overbar{F}_{i,n}^{\phi '} \underbar{$\Delta$}_{n}^{''} 
    - \epsilon \overbar{\Delta}_i^{\phi '} \overbar{H}_{i}^{\phi ''} + \epsilon^2 {\textstyle\sum\limits_{n \in \mathcal{N} }} \overbar{F}_{i,n}^{\Delta \phi '} \underbar{$\Delta$}_{n}^{''} +
    \overbar{E}_i^{\phi ''} \overbar{H}_{i}^{\phi '} -            \epsilon   {\textstyle\sum\limits_{n \in \mathcal{N} }} \overbar{F}_{i,n}^{\phi ''} \underbar{$\Delta$}_{n}^{'} 
    -\epsilon \overbar{E}_i^{\phi ''} \overbar{H}_{i}^{\phi '} +   \epsilon^2 {\textstyle\sum\limits_{n \in \mathcal{N} }} \overbar{F}_{i,n}^{\Delta \phi ''} \underbar{$\Delta$}_{n}^{'} \nonumber \\
    &\qquad \qquad \qquad \qquad \qquad \qquad \qquad \qquad \qquad \qquad \qquad \qquad \qquad \qquad \qquad \qquad \qquad \qquad \qquad \qquad \qquad \quad \ \hspace{2pt}  \forall i \in \mathcal{N}_{PQ} 
    \\
    & \lvert \overbar{E}_{i}^{II \phi} \rvert ^2 = (\overbar{E}_{i}^{\phi '})^2 - 2 \epsilon \overbar{E}_{i}^{\phi '} \overbar{\Delta}^{'}_{i} +  \epsilon^2 ( \overbar{\Delta}^{'}_{i})^2 + (\overbar{E}_{i}^{\phi ''})^2 - 2 \epsilon \overbar{E}_{i}^{\phi ''} \overbar{\Delta}^{''}_{i} +  \epsilon^2 ( \overbar{\Delta}^{''}_{i})^2 
    \qquad \qquad \qquad \qquad \qquad \quad \hspace{5pt}  \forall i \in \mathcal{N}_{PV} 
    \\
    &P^{II}_{j} =
    {E}_j {H}_{j} -               \epsilon   {\textstyle\sum\limits_{m \in \mathcal{M} }} {F}_{j,m} {\Delta}_{m} -
    \epsilon {\Delta}_j {H}_{j} + \epsilon^2 {\textstyle\sum\limits_{m \in \mathcal{M} }} {F}_{j,m}^{\Delta} {\Delta}_{m} \qquad \qquad \qquad \qquad \qquad \qquad \qquad \qquad \qquad \hspace{2pt} \forall j \in \mathcal{M}_{P}, 
    \\
    &P^{II}_{k} =
    {E}_k {H}_{k} -               \epsilon   {\textstyle\sum\limits_{m \in \mathcal{M} }} {F}_{k,m} {\Delta}_{m} -
    \epsilon {\Delta}_k {H}_{k} + \epsilon^2 {\textstyle\sum\limits_{m \in \mathcal{M} }} {F}_{k,m}^{\Delta} {\Delta}_{m} \qquad \qquad \qquad \qquad \qquad \qquad \qquad \qquad \quad \hspace{7pt} \forall k \in \mathcal{L}_{E_{dc}Q} 
    \\
    &P^{II,\phi}_{l} =
    \overbar{E}_l{\phi '} \overbar{H}_{l}^{\phi '} -                \epsilon   {\textstyle\sum\limits_{n \in \mathcal{N} }} \overbar{F}_{l,n}^{\phi '} \underbar{$\Delta$}_{n}^{'} -
    \epsilon \overbar{\Delta}_l^{\phi '} \overbar{H}_{l}^{\phi '} + \epsilon^2 {\textstyle\sum\limits_{n \in \mathcal{N} }} \overbar{F}_{l,n}^{\Delta \phi '} \underbar{$\Delta$}_{n}^{'}
    +\overbar{E}_l^{\phi ''} \overbar{H}_{l}^{\phi ''} +           \epsilon   {\textstyle\sum\limits_{n \in \mathcal{N} }} \overbar{F}_{l,n}^{\phi ''} \underbar{$\Delta$}_{n}^{''}
    +\epsilon \overbar{E}_l^{\phi ''} \overbar{H}_{l}^{\phi ''} -   \epsilon^2 {\textstyle\sum\limits_{n \in \mathcal{N} }} \overbar{F}_{l,n}^{\Delta \phi ''} \underbar{$\Delta$}_{n}^{''} \nonumber \\
    &\qquad \qquad \qquad \qquad \qquad \qquad   \forall l \in \mathcal{L}_{PQ}, 
    \\
    &Q^{II,\phi}_{l} =
    \overbar{E}_l^{\phi '} \overbar{H}_{l}^{\phi ''} -               \epsilon   {\textstyle\sum\limits_{n \in \mathcal{N} }} \overbar{F}_{l,n}^{\phi '} \underbar{$\Delta$}_{n}^{''} -
    \epsilon \overbar{\Delta}_l^{\phi '} \overbar{H}_{l}^{\phi ''} + \epsilon^2 {\textstyle\sum\limits_{n \in \mathcal{N} }} \overbar{F}_{l,n}^{\Delta \phi '} \underbar{$\Delta$}_{n}^{''} -
    \overbar{E}_l^{\phi ''} \overbar{H}_{l}^{\phi '} -               \epsilon   {\textstyle\sum\limits_{n \in \mathcal{N} }} \overbar{F}_{l,n}^{\phi ''} \underbar{$\Delta$}_{n}^{'} -
    \epsilon \overbar{E}_l^{\phi ''} \overbar{H}_{l}^{\phi '} +      \epsilon^2 {\textstyle\sum\limits_{n \in \mathcal{N} }} \overbar{F}_{l,n}^{\Delta \phi ''} \underbar{$\Delta$}_{n}^{'} \nonumber \\
    &\qquad \qquad \qquad \qquad \qquad \qquad \qquad \qquad \qquad \qquad \qquad \qquad \qquad \qquad \qquad \qquad \qquad \qquad \qquad \qquad \qquad \quad \ \hspace{2pt} \forall l \in \mathcal{L}_{E_{dc}Q}  \cup \mathcal{L}_{PQ} 
    \\
    & E_{k}^{II \ast} Y_{k,k}^{dc} E_{k}^{II \ast} +  {\textstyle\sum\limits_{\substack{m \in \mathcal{M} \setminus \{k\} }}} Y_{k,m}^{dc} E_{m} - \epsilon {\textstyle\sum\limits_{\substack{m \in \mathcal{M} \setminus \{k\} }}} Y_{k,m}^{dc} \Delta_{m} 
    =  \qquad \qquad \qquad \qquad \qquad \qquad \qquad \qquad \qquad  \forall (l,k) \in \mathcal{L}_{E_{dc}Q} \nonumber \\
    & \qquad \overbar{E}_l{\phi '} \overbar{H}_{l}^{\phi '} -                \epsilon   {\textstyle\sum\limits_{n \in \mathcal{N} }} \overbar{F}_{l,n}^{\phi '} \underbar{$\Delta$}_{n}^{'} -
    \epsilon \overbar{\Delta}_l^{\phi '} \overbar{H}_{l}^{\phi '} + \epsilon^2 {\textstyle\sum\limits_{n \in \mathcal{N} }} \overbar{F}_{l,n}^{\Delta \phi '} \underbar{$\Delta$}_{n}^{'}
    +\overbar{E}_l^{\phi ''} \overbar{H}_{l}^{\phi ''} +            \epsilon   {\textstyle\sum\limits_{n \in \mathcal{N} }} \overbar{F}_{l,n}^{\phi ''} \underbar{$\Delta$}_{n}^{''}
    +\epsilon \overbar{E}_l^{\phi ''} \overbar{H}_{l}^{\phi ''} -   \epsilon^2 {\textstyle\sum\limits_{n \in \mathcal{N} }} \overbar{F}_{l,n}^{\Delta \phi ''} \underbar{$\Delta$}_{n}^{''}\label{eq:proof_PF2_vdc}
\end{align}
\end{subequations}
\normalsize
\noindent
\end{figure*}

\noindent
Next, we subtract the PF models in \eqref{eq:proof_PF1} and \eqref{eq:proof_PF2} for network \textit{I} and \textit{II} to obtain \eqref{eq:proof_sol}. By \eqref{eq:proof_sol}, it follows that 
$P^{I,\phi}_i = P^{II,\phi}_i \ \forall i \in \mathcal{N}_{PQ} \cup \mathcal{N}_{PV}$, 
$Q^{I,\phi}_i = Q^{II,\phi}_i \ \forall i \in \mathcal{N}_{PQ}$, 
$P^{I}_j = P^{II}_j \ \forall j \in \mathcal{M}_{P}$, 
$P^{I}_k = P^{II}_k \ \forall k \in \mathcal{L}_{PQ}$,
$P^{I,\phi}_l = P^{II,\phi}_l \ \forall l \in \mathcal{L}_{PQ} $, 
$Q^{I,\phi}_l = Q^{II,\phi}_l \ \forall l \in \mathcal{L}_{PQ} \cup \mathcal{L}_{V_{dc}Q}$,  
and 
$\lvert \overbar{E}_{i}^{I \phi} \rvert = \lvert \overbar{E}_{i}^{II \phi} \rvert  \ \forall i \in \mathcal{N}_{PV} $, 
$E^{I}_j = E^{II}_j \ \forall j \in \mathcal{M}_{V}$, 
$E^{I}_k = E^{II}_k \ \forall k \in \mathcal{L}_{V_{dc}Q}$.
Therefore, network \textit{ I} and \textit{II} have the same power injections in the power-controllable nodes and the same voltages at the voltage-controllable nodes. According to hypothesis of \textbf{Corollary 1}, the PF Jacobian, i.e. the derivative of the unified PF equations, is invertible. Next, we apply the inverse function theorem that states that the PF equations are also invertible in a neighbourhood around the current operating point. Now we take an $\epsilon$ arbitrary small so all the voltages $\overbar{E}^{I}$ and $\overbar{E}^{II}$ belong to this neighbourhood with a one-to-one mapping between the voltages and the power injections. Because we showed before in \eqref{eq:proof_sol} that the active and reactive power injections are exactly the same in network \textit{I} and network \textit{II}, the voltages $\overbar{E}^{I}$ and $\overbar{E}^{II}$ must also be exactly the same. Therefore, \eqref{eq:proof_V1} = \eqref{eq:proof_V2}, and thus $\overbar{\Delta}_i, {\Delta}_j,\overbar{\Delta}_l,{\Delta}_k = 0$ for all $i\in \mathcal{N}, j\in \mathcal{M}, (l,k)\in \mathcal{L} $.
The homogeneous system of equations in \eqref{eq:proof_homo} only has the trivial solution. Therefore, the uniqueness of the solution is proved.

\small
\noindent
\begin{subequations} \label{eq:proof_sol}
\begin{align} 
    &P^{I,\phi}_{i} - P^{II,\phi}_{i}  = \qquad \qquad \qquad \qquad \qquad \qquad  \forall i \in \mathcal{N}_{PQ} \cup \mathcal{N}_{PV} \nonumber \\
    &  \ \ 2 \epsilon \Big( 
    \overbar{H}_{i}^{\phi '}  \overbar{\Delta}^{\prime}_{i}
     +  {\textstyle\sum\limits_{n \in \mathcal{N} }} \overbar{F}_{i,n}^{\phi '}  \overbar{\Delta}^{\prime}_{n}  - \overbar{H}_{i}^{\phi ''} \overbar{\Delta}^{\prime \prime}_{i}
    +  {\textstyle\sum\limits_{n \in \mathcal{N} }} \overbar{F}_{i,n}^{\phi ''}  \overbar{\Delta}^{\prime \prime}_{n}  \Big)  \\
    &Q^{I,\phi}_{i} - Q^{II,\phi}_{i}  = \qquad \qquad \qquad \qquad \qquad \qquad  \forall i \in \mathcal{N}_{PQ} \nonumber \\
    &  \ \ 2 \epsilon \Big( 
    \overbar{H}_{i}^{\phi ''} \overbar{\Delta}^{\prime}_{i}
    +  {\textstyle\sum\limits_{n \in \mathcal{N} }} \overbar{F}_{i,n}^{\phi '}  \overbar{\Delta}^{\prime}_{n}   + \overbar{H}_{i}^{\phi '}  \overbar{\Delta}^{\prime \prime }_{i}
    -  {\textstyle\sum\limits_{n \in \mathcal{N} }} \overbar{F}_{i,n}^{\phi ''}  \overbar{\Delta}^{\prime \prime}_{n} \Big)  \\
    & \lvert \overbar{E}_{i}^{I \phi} \rvert ^2 - \lvert \overbar{E}_{i}^{II \phi} \rvert ^2  = 
    2 \epsilon \Big(  \overbar{E}_{i}^{\phi '} \overbar{\Delta}^{\prime}_{i} +  \overbar{E}_{i}^{\phi ''}\overbar{\Delta}^{\prime \prime}_{i},  \Big), \forall i \in \mathcal{N}_{PV} 
    \\
    &P^{I}_{j} - P^{II}_{j}  = 2 \epsilon \Big( 
     {H}_{j}  \overbar{\Delta}_{j}
     +  {\textstyle\sum\limits_{m \in \mathcal{M} }} {F}_{j,m}  {\Delta}_{m} \Big), \quad \forall j \in \mathcal{M}_{P} \\
    & P^{I}_{k} - P^{II}_{k} = 2 \epsilon \Big( 
    {H}_{k} {\Delta}_{k} +  {\textstyle\sum\limits_{m \in \mathcal{M} }} {F}_{k,m}  {\Delta}_{m} \Big), \quad  \forall k \in \mathcal{L}_{PQ} 
    \\
    & P^{I,\phi}_{l} - P^{II,\phi}_{l} = \qquad \qquad \qquad \qquad \qquad \qquad \forall l \in \mathcal{L}_{PQ} \nonumber\\
    & \ \ 2 \epsilon \Big( 
       \overbar{{H}}_{i}^{\phi '}  {\overbar{\Delta}}^{'}_{l}
     +  {\textstyle\sum\limits_{n \in \mathcal{N} }} \overbar{{F}}_{l,n}^{\phi '} {\overbar{\Delta}}^{'}_{n}
     - \overbar{{H}}_{i}^{ \phi ''}  {\overbar{\Delta}}^{''}_{l}  
     +  {\textstyle\sum\limits_{n \in \mathcal{N} }} \overbar{{F}}_{l,n}^{\phi ''}  {\overbar{\Delta}}^{''}_{n}  \Big) 
     \\
    & Q^{I,\phi}_{l} - Q^{II,\phi}_{l} = \qquad \qquad \qquad \qquad \qquad \qquad \forall l \in \mathcal{L}_{E_{dc}Q}  \cup \mathcal{L}_{PQ} \nonumber \\
    & \ \  2 \epsilon \Big( 
      \overbar{{H}}_{i}^{\phi ''} {\overbar{\Delta}}^{'}_{l} +  {\textstyle\sum\limits_{n \in \mathcal{N} }} \overbar{{F}}_{l,n}^{\phi ''} {\overbar{\Delta}}^{'}_{n}
      + \overbar{{H}}_{i}^{\phi '}   {\overbar{\Delta}}^{''}_{l}  -   {\textstyle\sum\limits_{n \in \mathcal{N} }} \overbar{{F}}_{l,n}^{\phi '}  {\overbar{\Delta}}^{''}_{n}  \Big)  
    \\
    & \eqref{eq:proof_PF1_vdc} - \eqref{eq:proof_PF2_vdc}  =  \qquad \qquad  \qquad \qquad \qquad \qquad \forall (l,k) \in \mathcal{L}_{E_{dc}Q} \nonumber \\
    & \ \  2 \epsilon \Big( 
     \overbar{{H}}_{i}^{\phi '} {\overbar{\Delta}}^{'}_{l}
     +  {\textstyle\sum\limits_{n \in \mathcal{N} }} \overbar{{F}}_{l,n}^{\phi '}  {\overbar{\Delta}}^{'}_{n}   - \overbar{{H}}_{i}^{\phi ''}  {\overbar{\Delta}}^{''}_{l} +  {\textstyle\sum\limits_{n \in \mathcal{N} }} \overbar{{F}}_{l,n}^{\phi ''}  {\overbar{\Delta}}^{''}_{n} \nonumber \\
     &  - {\textstyle\sum\limits_{m \in \mathcal{M} \setminus \{k\} }}  F_{k,m} {\Delta}_{m} \Big) 
\end{align}
\end{subequations}
\normalsize

\bibliographystyle{IEEEtran}
\bibliography{References}

\end{document}